\documentclass[journal]{IEEEtran}

\usepackage{verbatim}
\usepackage{amsfonts}
\usepackage{amssymb}
\usepackage{stfloats}
\usepackage{cite}
\usepackage{graphicx}
\usepackage{psfrag}
\usepackage{subfigure}
\usepackage{amsmath}
\usepackage{array}
\usepackage{epstopdf}
\usepackage{authblk}
\usepackage{graphicx} 
\usepackage{amsthm} 
\usepackage{lipsum}
\usepackage{verbatim} 
\usepackage{authblk}
\usepackage{mathtools}
\usepackage{cuted}
\usepackage{booktabs}
\usepackage{cases}
\usepackage{multirow}
\usepackage{bbding}
\usepackage{pifont}
\usepackage{makecell}
\usepackage{threeparttable}
\usepackage[bookmarks=true, colorlinks=true, breaklinks=true]{hyperref}

\newtheorem{theorem}{Theorem}

\newtheorem{lemma}{Lemma}
\newtheorem{corollary}{Corollary}

\newtheorem{assumption}{Assumption}

\newtheorem{remark}{\bf Remark}

\def\phi{\varphi}

\def\({\left(}
\def\){\right)}

\def\b0{{\mathbf{0}}}

\usepackage{algorithmic}
\usepackage{algorithm}

\usepackage{setspace}

\usepackage{xcolor}

\usepackage{epstopdf}

\makeatletter

\newcommand{\Rmnum}[1]{\expandafter\@slowromancap\romannumeral #1@}
\makeatother

\begin{document}
	\bstctlcite{ref:BSTcontrol}
	
	\title{Semi-Federated Learning: Convergence Analysis and Optimization of A Hybrid Learning Framework}
	
	\author{Jingheng~Zheng,~Wanli~Ni,~Hui~Tian,~\IEEEmembership{Senior~Member,~IEEE,}~Deniz~Gündüz,~\IEEEmembership{Fellow,~IEEE,}
		Tony~Q.~S.~Quek,~\IEEEmembership{Fellow,~IEEE,}~and~Zhu~Han,~\IEEEmembership{Fellow,~IEEE}
		\vspace{-0.4 cm}
		\thanks{The work of Hui Tian was supported by the National Natural Science Foundation of China under Grant 62071068. The work of Tony Q. S. Quek was supported by the National Research Foundation, Singapore and Infocomm Media Development Authority under its Future Communications Research \& Development Programme. The work of Zhu Han was supported by NSF CNS-2107216, CNS-2128368, CMMI-2222810, US Department of Transportation, Toyota and Amazon. The work of Jingheng Zheng was supported by the China Scholarship Council.
		This work was presented in part at the IEEE International Conference on Computer Communication (INFOCOM) Workshops, New York, NY, USA, May 2022~\cite{Zheng2022Semi}.~\textit{(Corresponding author: Hui Tian.)}}
		\thanks{J.~Zheng,~W.~Ni and~H.~Tian are with the State Key Laboratory of Networking and Switching Technology, Beijing University of Posts and Telecommunications, Beijing 100876, China (e-mail: zhengjh@bupt.edu.cn; charleswall@bupt.edu.cn; tianhui@bupt.edu.cn).}
		\thanks{D.~Gündüz is with the Department of Electrical and Electronic Engineering, Imperial College London, London SW7 2AZ, UK, and also with the Department of Engineering ``Enzo Ferrari," University of Modena and Reggio Emilia (Unimore), Modena 41100, Italy (e-mail: d.gunduz@imperial.ac.uk).}
		\thanks{T.~Q.~S.~Quek is with the Singapore University of Technology and Design, Singapore 487372, and also with the Department of Electronic Engineering, Kyung Hee University, Yongin 17104, South Korea (e-mail: tonyquek@sutd.edu.sg).}
		\thanks{Z.~Han is with the Department of Electrical and Computer Engineering, University of Houston, Houston, TX 77004, USA, and also with the Department of Computer Science and Engineering, Kyung Hee University, Seoul 446-701, South Korea (e-mail: hanzhu22@gmail.com).}
	}
	\maketitle

\begin{abstract}
	Under the organization of the base station (BS), wireless federated learning (FL) enables collaborative model training among multiple devices.
	However, the BS is merely responsible for aggregating local updates during the training process, which incurs a waste of the computational resource at the BS.
	To tackle this issue, we propose a semi-federated learning (SemiFL) paradigm to leverage the computing capabilities of both the BS and devices for a hybrid implementation of centralized learning (CL) and FL.
	Specifically, each device sends both local gradients and data samples to the BS for training a shared global model.
	To improve communication efficiency over the same time-frequency resources, we integrate over-the-air computation for aggregation and non-orthogonal multiple access for transmission by designing a novel transceiver structure.
	To gain deep insights, we conduct convergence analysis by deriving a closed-form optimality gap for SemiFL and extend the result to two extra cases.
	In the first case, the BS uses all accumulated data samples to calculate the CL gradient, while a decreasing learning rate is adopted in the second case.
	Our analytical results capture the destructive effect of wireless communication and show that both FL and CL are special cases of SemiFL.
	Then, we formulate a non-convex problem to reduce the optimality gap by jointly optimizing the transmit power and receive beamformers.
	Accordingly, we propose a two-stage algorithm to solve this intractable problem, in which we provide the closed-form solutions to the beamformers.
	Extensive simulation results on two real-world datasets corroborate our theoretical analysis, and show that the proposed SemiFL outperforms conventional FL and achieves $3.2\%$ accuracy gain on the MNIST dataset compared to state-of-the-art benchmarks.
\end{abstract}

\begin{IEEEkeywords}
	Semi-federated learning, communication efficiency, convergence analysis, transceiver design.
\end{IEEEkeywords}
	
\section{Introduction}
As a thriving distributed learning framework, wireless federated learning (FL) enables multiple clients (e.g., devices) to collaboratively train a shared model by iteratively exchanging their local updates (e.g., model parameters or gradients) with the parameter server (e.g., the base station (BS))~\cite{Chen2021Distributed,Wahab2021Federated,Khan2020Federated}.
Compared to centralized learning (CL), FL features data privacy preservation, reduced communication cost, and fast inference~\cite{Lim2020Federated,Ye2020Federated}.
However, in the conventional FL paradigm, only the distributed computational resources of local devices are utilized to complete the model training~\cite{Ruby2022Energy}.
The powerful computing capability at the BS is insufficiently involved in the learning task.
This raises an intuitive problem: \textit{how to exploit the underutilized computational resource at the BS to improve the performance of FL}?
To solve this problem, one potential strategy is to provide the BS with some data samples from devices so that its computational resources can be utilized to further promote the model performance~\cite{Ni2023Semi,Elbir2021Hybrid}.

Apart from how to compute model, another critic issue of FL is how to transmit local updates in wireless networks.
In the literature, there are three commonly adopted wireless communication schemes for transmitting local updates from devices to the BS, including orthogonal multiple access (OMA)~\cite{Chen2021A}, non-orthogonal multiple access (NOMA)~\cite{Ni2022Integrating,Wu2022Non}, and over-the-air computation (AirComp)~\cite{Zheng2022Balancing,Yang2020Federated,Zhu2021One}.
Specifically, in OMA-based FL schemes, each device occupies a dedicated resource block in the time or frequency domain to avoid interference.
However, the transmission bandwidth or time of OMA decreases with the number of devices, which results in a higher communication latency and thus taking more time to reach the global convergence.
By allowing all devices to share the resource block, NOMA-based FL schemes are beneficial to support massive connectivity and improve the throughput~\cite{Zeng2017Capacity}, thus accelerating the training speed.
However, the co-channel interference introduced by NOMA brings new challenges for the transceiver design and signal decoding.
For instance, if the transmit power of local devices is improperly allocated, it will increase the inter-device interference when the BS decodes individual signals, and even prevent the BS from decoding the signals correctly, thereby reducing the achievable data rate of uplink transmissions.

Note that both OMA- and NOMA-based schemes are communication-centric paradigms, which do not directly match the model aggregation process of FL tasks.
Recently, by aggregating local updates based on the superposition property of the wireless channel~\cite{Yang2020Federated}, AirComp-based FL schemes have drawn considerable attention for their unique advantages, such as exploiting interference for computation and reducing latency.
As mentioned previously, in order to make full use of the computing resources of the BS, local devices can upload some data samples to the BS while uploading the local updates.
However, AirComp-based schemes mainly focus on function computation instead of decoding individual data streams, which makes it impossible for devices to simultaneously upload data samples and local updates in a spectrum-efficient manner.
This raises a challenging problem: \textit{how to design a spectral-efficient joint communication and computation (JCC) scheme that supports the concurrent uplink transmission of data samples and local updates using the shared time-frequency resources}?
Intuitively, one can directly incorporate communication-efficient NOMA and computation-efficient AirComp into a harmonized multiple access scheme so that their respective advantages can be sufficiently leveraged.
Nevertheless, the transceiver structure for supporting such an integrated scheme needs to be meticulously designed to mitigate the severe co-channel interference.

When addressing the aforementioned problems, we encounter the following challenges.
First, although utilizing the computing capability of the BS for model training is expected to accelerate the convergence, the involvement of the BS complicates the convergence analysis of the training process due to the need of rigorous mathematical knowledge and skillful derivations~\cite{Cao2022Transmission,Chen2021A}.
Second, since both the datasets and local updates are transmitted over the non-ideal wireless channels, their impacts on the convergence should also be quantified precisely~\cite{Wan2021Convergence}.
Third, existing resource allocation schemes designed for conventional communication of FL systems are not applicable to the collaborative learning framework requiring the concurrent transmission of data samples and local updates.
Therefore, it is imperative to develop new transceiver control algorithms that can further improve both the communication efficiency and learning performance of the considered new learning framework.

\subsection{Contributions and Organization}
To leverage the underutilized computing resources at the BS for improving learning performance, we propose a novel semi-federated learning (SemiFL) paradigm by integrating the conventional CL and FL into a two-tier framework.
Since many previous studies on FL solely focus on transmitting local gradients, the existing communication strategies can not directly satisfy the requirement of SemiFL for collecting both local gradients and data samples.
To address this issue, we propose a novel JCC scheme to guarantee the unique communication request of SemiFL in an efficient manner.
Specifically, at the devices, we combine AirComp and NOMA techniques to enable the concurrent transmission of data samples and local gradients.
To gain deep insights into SemiFL, we provide the theoretical analysis of SemiFL.
Then, we formulate an optimality gap minimization problem by optimizing the transceivers.
The main contributions of this paper are summarized as follows:
\begin{itemize}
	\item To improve the learning performance of existing FL, we propose a harmonized SemiFL framework to orchestrates CL and FL into a two-tier architecture.
 	The global model at the BS is updated by the hybrid gradient obtained from both CL and FL.
	Different from conventional communication strategies for FL, we design a learning-centric JCC transceiver structure to meet the unique transmission requirement of SemiFL.
	At the transmitter, the devices concurrently transmit local gradients and data samples via a shared multiple access channel.
	At the receiver, the BS first decodes the data samples for CL, and then aggregates the local gradients over the air.
	\item We derive the optimality gap in closed form to characterize the impact of wireless communication on the convergence performance of SemiFL.
	We	further extend the result to two special cases, where the BS calculates the CL gradient using all accumulated data sample in the first case and adopts a decreasing learning rate in the second case.
	By comparing the learning behaviors between SemiFL, FL, and CL, we theoretically prove that SemiFL is a more general learning paradigm than the other two.
	To further accelerate the convergence of SemiFL, we formulate a non-convex problem to minimize the optimality gap by designing the transceivers while satisfying the maximum transmit power of devices, the communication latency of data transmission, and the distortion of gradient aggregation.
	\item We propose a two-stage algorithm to solve the formulated challenging problem.
	Specifically, we provide the closed-form solution to the aggregation beamformer in the single-antenna case.
	Moreover, the successive convex approximation (SCA) method is employed to obtain the decoding beamformers, where the closed-form optimal solutions in each iteration are derived by solving the Karush-Kuhn-Tucker (KKT) conditions.
\end{itemize}

Apart from the contributions, simulation results on two real-world datasets confirm that:
\begin{enumerate}
	\item The proposed two-stage algorithm outperforms benchmarks in terms of aggregation mean square error (MSE) and communication sum rate.
	\item The proposed SemiFL achieves higher learning accuracy and faster convergence than FL, which validates the theoretical relation between SemiFL, FL, and CL.
	\item Compared with state-of-the-art benchmarks, SemiFL achieves up to $3.2\%$ accuracy gain on the MNIST dataset, and the effectiveness of the proposed two-stage resource allocation algorithm is validated.
\end{enumerate}

The rest of this paper is organized as follows.
The system model of SemiFL is described in Section~\ref{system_model}.
Section~\ref{convergence_analysis_and_problem_formulation} presents the convergence analysis and formulates the problem.
Section~\ref{optimization_of_the_transmit_power_control_policies_and_receive_beamformers} proposes the two-stage algorithm to jointly optimize the transmit power and receive beamformers.
Section~\ref{simulation_results} presents simulation results, followed by conclusions in Section~\ref{conclusion}.

\emph{Notations}:~Lower-case and upper-case boldface letters denote vectors and matrices, respectively.
Lower-case letters and upper-case cursive letters denote scalars and sets, respectively.
${\bf{I}}_N$ denotes the $N \times N$ identity matrix.
$|\cdot|$ denotes the cardinality of a set or the modulus of a complex scalar.
$(\cdot)^{\rm T}$, $(\cdot)^{\rm H}$, and $\| \cdot \|$ denote transpose, conjugate transpose, and vector 2-norm, respectively.
$\mathbb{R}$, $\mathbb{C}$, and $\varnothing$ denote real, complex, and empty sets, respectively.
$\nabla$ denotes the gradient operator and $\mathbb{E}[\cdot]$ takes statistical expectation.
$\cup$ and $\cap$ denote the union and intersection of sets, respectively.
${\log}_2 (\cdot)$ takes the base two logarithm, and $\lim\limits_{x \rightarrow \infty}{f(x)}$ denotes the limit of $f(x)$ as $x$ approaches infinity.
${\mathop{\rm Re}\nolimits}\{x\}$ and $\angle x$ denote the real part and the angle of a complex scalar $x$, respectively.
	
\begin{figure}
	\centering
	\includegraphics[width=0.48\textwidth]{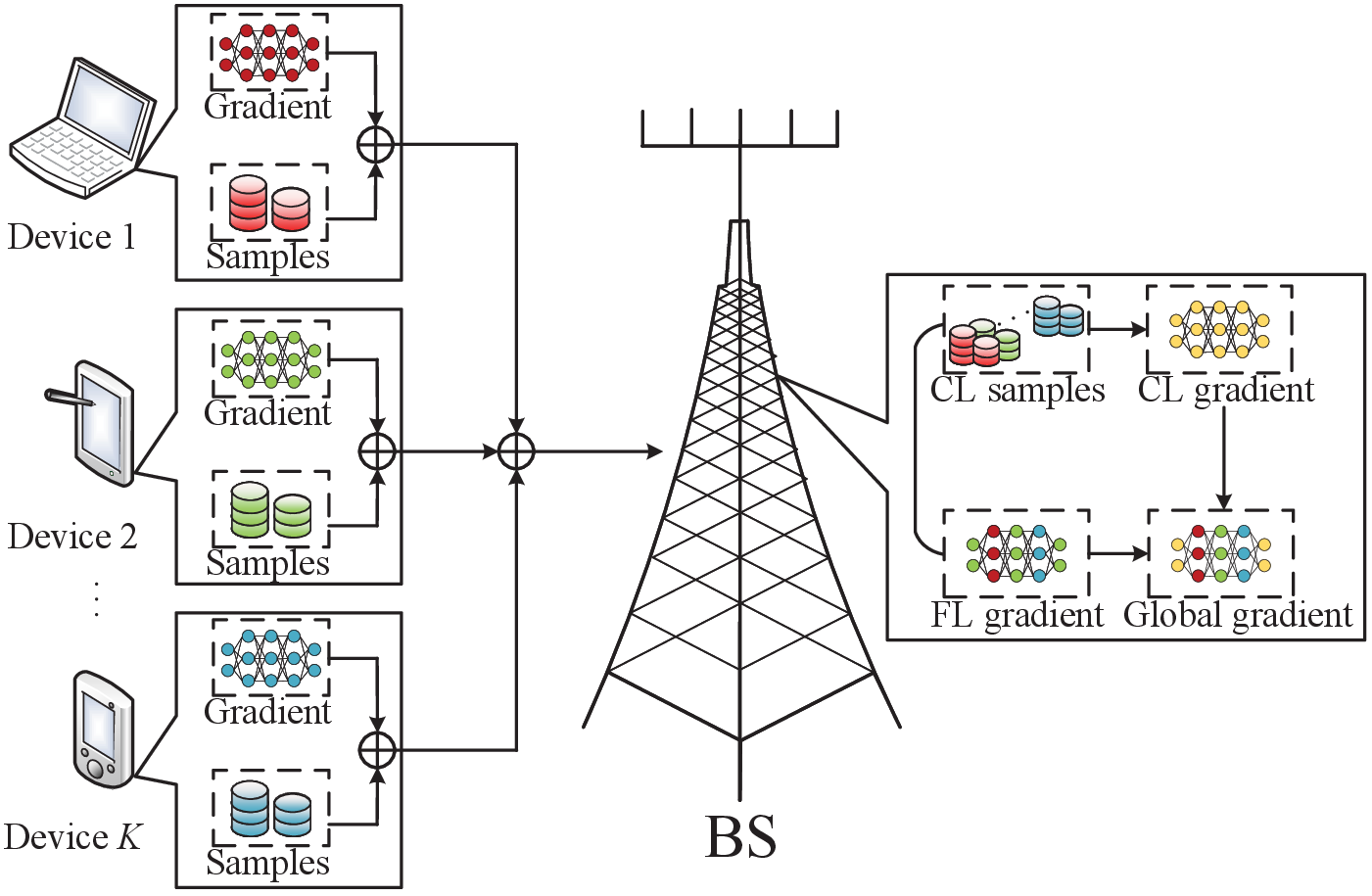}
	\caption{A demonstration of the proposed SemiFL framework.}
	\label{system_model_figure}
	\vspace{-0.4 cm}
\end{figure}

\section{System Model}
\label{system_model}
As depicted in Fig.~\ref{system_model_figure}, we consider a wireless network comprising one $N_r$-antenna BS and  $K$ single-antenna devices. 
Specifically, all devices that collect data samples and conduct local computation form the first tier, while the BS performing centralized computation serves as the second tier.
The set of devices is denoted by $\mathcal{K}=\{1,2,\ldots,K\}$.

\subsection{SemiFL Framework}
\label{federated_leanring_model}
We consider a model training process with $T$ communication rounds.
In the $t$-th round, the $k$-th device collects multiple data samples denoted by a dataset $\mathcal{D}_{t,k}$, which is divided into two disjoint subsets, i.e., $\mathcal{D}_{f,t,k}$ containing $N_{f,k}$ samples and $\mathcal{D}_{c,t,k}$ containing $N_{c,k}$ samples, satisfying $\mathcal{D}_{f,t,k} \cup \mathcal{D}_{c,t,k}=\mathcal{D}_{t,k}$ and $\mathcal{D}_{f,t,k} \cap \mathcal{D}_{c,t,k}=\varnothing$.
Note that $|\mathcal{D}_{t,k}|=N_{f,k}+N_{c,k}$.
Denote $\mathcal{D}_k={\cup}_{t=1}^{T}{\mathcal{D}_{k,t}}$ as the dataset that encompasses all data samples collected by the $k$-th device over $T$ rounds.
Local devices aim to collaboratively train a shared global model ${\bf{w}} \in \mathbb{R}^{Q}$ by minimizing the global empirical loss function $F({\bf{w}})$ on the global dataset $\mathcal{D}={\cup}_k \mathcal{D}_k$, which is given by 
\begin{align}
	F({\bf{w}})=\frac{1}{N} \sum \nolimits_{k=1}^{K} \sum \nolimits_{n \in \mathcal{D}_k} {f({\bf{w}};{\bf{x}}_{k,n},{\bf{y}}_{k,n})},
\end{align}
where ${\bf{x}}_{k,n}$ and ${\bf{y}}_{k,n}$ are the feature vector and the label vector of a data sample, respectively, $f({\bf{w}};{\bf{x}}_{k,n},{\bf{y}}_{k,n})$ is the loss function with respect to (w.r.t.) a data sample, and $N=|\mathcal{D}|=\sum\nolimits_{k=1}^K \sum\nolimits_{t=1}^{T} (N_{f,k}+N_{c,k})$ is the total amount of data samples collected by all $K$ devices over $T$ rounds.
Different from conventional FL where the global model $\bf{w}$ is merely updated by the aggregated local gradients, we propose a SemiFL framework to minimize the global empirical loss function $F({\bf{w}})$. 
Specifically, FL over devices and CL of the BS are coordinated in a unified manner, and the global model is updated by a combination of the resultant FL gradient and CL gradient.

In the $t$-th round, limited by the local computing capability, the $k$-th device calculates the local gradient ${\bf{g}}^f_{t,k}\!\!=\!\![g^f_{t,k,1},\ldots,g^f_{t,k,Q}]^{\rm T} \!\! \in \! \mathbb{R}^{Q}$ using the $N_{f,k}$ data samples in $\mathcal{D}_{f,t,k}$, given by
\begin{align}
	{\bf{g}}^f_{t,k}=\frac{1}{N_{f,k}} \sum\nolimits_{n \in \mathcal{D}_{f,t,k}}{{\bf{g}}_{t,k,n}},~\forall k \in \mathcal{K},
\end{align}
where ${\bf{g}}_{t,k,n}\triangleq{\nabla}f({\bf{w}}_t;{\bf{x}}_{k,n},{\bf{y}}_{k,n})$ is the sample-wise gradient at ${\bf{w}}_t$, and ${\bf{w}}_t$ denotes the global model in the $t$-th round.
Note that the privacy of the data in $\mathcal{D}_{f,t,k}$ can be preserved naturally since the BS has no access to the raw data retained by local devices.
Apart from transmitting ${\bf{g}}^{f}_{t,k}$ for aggregation, the $k$-th device also uploads the data samples in $\mathcal{D}_{c,t,k}$ to the BS.
To mitigate privacy leakage, we employ a mixup method, originally proposed in~\cite{Zhang2018Mixup}, to preserving privacy when sending data to third parties~\cite{Oh2020Mix2FLD,Koda2021AirMixML,Park2021Communication}.
For an arbitrary sample $\{{\bf{x}}_{t,k,n},{\bf{y}}_{t,k,n}\}$ in $\mathcal{D}_{c,t,k}$, the $k$-th device mixes it with another sample labeled differently using a mixed ratio $\varpi \in (0,1)$ drawn from a Dirichlet distribution, and then adds noise to the mixed sample to enhance privacy preservation.
Specifically, the mixed data sample $\{\bar{\bf{x}}_{t,k,n},\bar{\bf{y}}_{t,k,n}\}$ is generated by
\begin{align}
	\bar{\bf{x}}_{t,k,n}=&\varpi{\bf{x}}_{t,k,n}+(1-\varpi){\bf{x}}_{t,k,n^{'}} \notag \\
	&+\bar{\bf{n}}_{t,k,n},\forall k \in \mathcal{K}, \forall n,n^{'} \in \mathcal{D}_{c,t,k}, \\
	\bar{\bf{y}}_{t,k,n}=&\varpi{\bf{y}}_{t,k,n}+(1-\varpi){\bf{y}}_{t,k,n^{'}} \notag \\
	&+\bar{\bf{n}}_{t,k,n},\forall k \in \mathcal{K}, \forall n,n^{'} \in \mathcal{D}_{c,t,k},
\end{align}
where ${\bf{y}}_{t,k,n} \neq {\bf{y}}_{t,k,n^{'}}$, and $\bar{\bf{n}}_{t,k,n}$ denotes the Gaussian noise whose strength can be adjusted to achieve a specific privacy level~\cite{Koda2021AirMixML}.
Namely, the $k$-th device sends the mixed data samples with noise to the BS so that the data privacy can be preserved. 

The BS accumulates all mixed data samples received, and randomly selects $\!N_c\!=\!\sum\nolimits_{k=1}^K {N_{c,k}}$ samples to form a dataset $\mathcal{D}_{c,t}$ for calculating the CL gradient, given by
\vspace{-0.2 cm}
\begin{align}
	\label{centralized_learning}
	{\bf{g}}_t^c=\frac{1}{N_c} \sum \nolimits_{n \in \mathcal{D}_{c,t}} {{\bf{g}}_{t,n}}.
\end{align}
Then, the BS aggregates the local gradients over the air.
Let ${\hat{\bf{g}}_t^f}\in\mathbb{R}^Q$ denote the aggregated gradient of FL.
Next, the BS calculates the global gradient $\hat{\bf{g}}_t$ for the $t$-th round as a combination of the FL gradient and CL gradient, i.e., a weighted average of $\hat{\bf{g}}_t^f$ and ${\bf{g}}_t^c$, given by
\begin{align}
	\label{model_combine_ideal}
	\hat{\bf{g}}_{t}= \frac{N_f}{N_f+N_c} \hat{\bf{g}}_t^f + \frac{N_c}{N_f+N_c} {\bf{g}}_t^c,
\end{align}
where $N_f\!=\!\sum\nolimits_{k=1}^K {N_{f,k}}$ is the total amount of data for FL in the $t$-th round.
Finally, the BS updates the global model ${\bf{w}}_{t}$ for the next round using the gradient descent method, i.e., ${\bf{w}}_{t+1}={\bf{w}}_{t} - \eta \hat{\bf{g}}_{t}$, where $\eta$ is the learning rate.

\begin{figure*}
	\centering
	\includegraphics[width=1\textwidth]{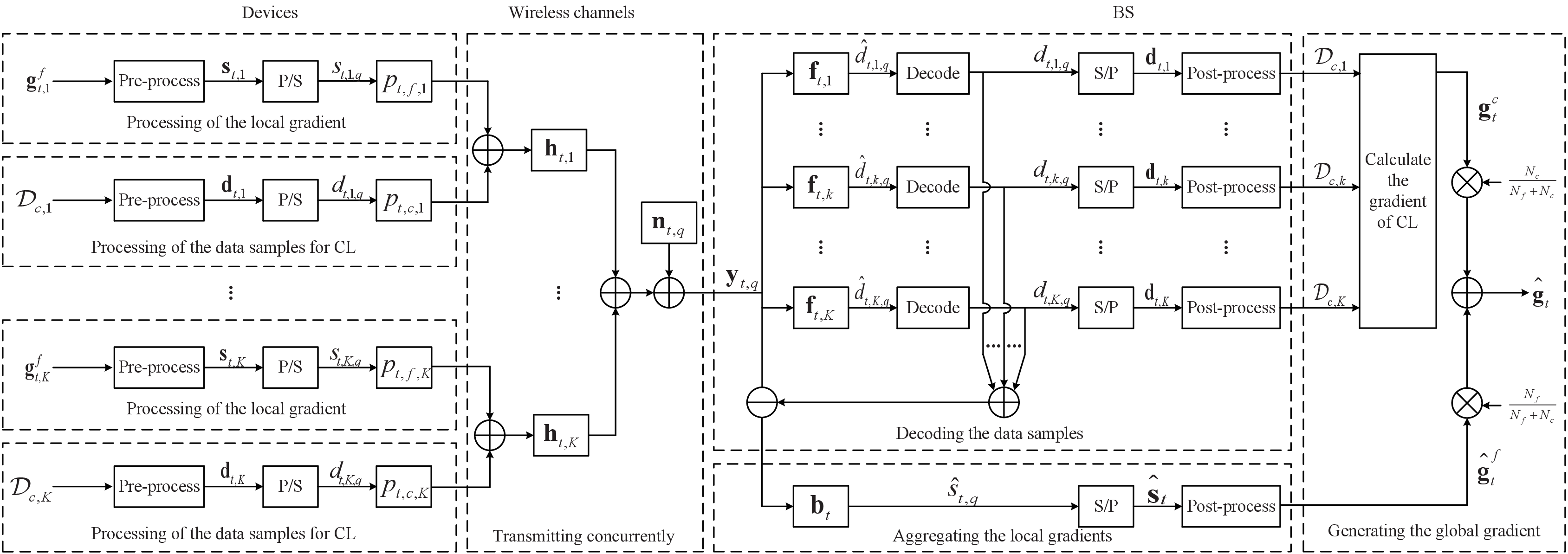}
	\caption{The block diagram of the designed JCC transceiver structure and signal processing flows for the SemiFL framework.}
	\label{block_diagram}
\end{figure*}

\subsection{JCC Scheme}
\label{communication_model}
In order to meet the unique transmission requirement of SemiFL, i.e., the collaborative transmission of local gradients and data samples, we propose a JCC scheme which simultaneously implements AirComp and NOMA in a communication-efficient manner. 
To this end, we design a novel transceiver structure for supporting the combination of these two critical techniques, as illustrated in Fig.~\ref{block_diagram}.
Specifically, the devices transmit both local gradients and data samples over the same time-frequency resources, while the BS first decodes the data samples, and then aggregates the local gradients over the air.

In the $t$-th communication round, the signal pre-processing of the $k$-th device is two-fold.
Here, the local gradient ${\bf{g}}^f_{t,k}$ is first normalized to a vector ${\tilde{\bf{g}}}^f_{t,k}=[{\tilde{g}}^f_{t,k,1},\ldots,{\tilde{g}}^f_{t,k,Q}]^{\rm T} \in \mathbb{R}^{Q}$ yielding $\mathbb{E}[{\tilde{g}}^f_{t,k,q}]\!=\!0$ and $\mathbb{E}[({\tilde{g}}^f_{t,k,q})^2]\!=\!1, \forall k \in \mathcal{K}$, and then transformed to a signal vector ${\bf{s}}_{t,k}=[s_{t,k,1},\ldots,s_{t,k,Q}]^{\rm T} \in \mathbb{R}^{Q}$.
Concretely, similar to~\cite{Zhu2020Broadband}, the normalization procedure is illustrated as follows:
\begin{enumerate}
	\item Before the transmission of local gradients and data samples, the $k$-th device calculates and transmits two parameters to the BS, i.e., $\frac{1}{Q}\sum_{q=1}^Q {g_{t,k,q}^f}$ and $\frac{1}{Q}\sum_{k=1}^K {(g_{t,k,q}^f)^2}$, where $g_{t,k,q}^f$ denotes the $q$-th entry of the local gradient ${\bf{g}}^f_{t,k}$.
	\item Upon receiving all $2K$ parameters uploaded by $K$ devices, the BS calculates the global mean $\bar{g}_t$ and the global variance $\bar{\sigma}_t^2$ by using $\bar{g}_t=\frac{1}{K} \sum_{k=1}^K (\frac{1}{Q}\sum_{q=1}^Q {g_{t,k,q}^f})$ and $\bar{\sigma}_t^2=\frac{1}{K}\sum_{k=1}^K [\frac{1}{Q} \sum_{q=1}^Q (g_{t,k,q}^f)^2] - \bar{g}_t^2$, respectively.
	\item The BS stores $\bar{g}_t$ and $\bar{\sigma}_t^2$ for the de-normalization in the post-processing, and broadcasts $\bar{g}_t$ and $\bar{\sigma}_t^2$ back to all devices for the normalization in the pre-processing.
	\item The $k$-th device normalizes the $q$-th entry of the local gradient ${\bf{g}}^f_{t,k}$ according to
	\begin{align}
		\label{normalization}
		\tilde{g}^f_{t,k,q}= \frac{g^f_{t,k,q}\! -\! \bar{g}_t}{\bar{\sigma}_t},~ q=1,2,\ldots,Q ,\ \forall k \in \mathcal{K},
	\end{align}
	where the normalized ${\tilde{g}}^f_{t,k,q}$ yields $\mathbb{E}[{\tilde{g}}^f_{t,k,q}]\!=\!0$ and $\mathbb{E}[({\tilde{g}}^f_{t,k,q})^2]\!=\!1$. Then, the $k$-th device constructs the gradient signal vector as ${\bf{s}}_{t,k}=\frac{N_{f,k}}{N_f}{\tilde{\bf{g}}}^f_{t,k}$.
\end{enumerate}
Note that the communication overhead of the normalization is $2K$ parameters in each round.
For another, the $N_{c,k}$ data samples in $\mathcal{D}_{c,t,k}$ for uploading, represented in bits, generally have different dimensions compared to the local gradient.
However, the signal vector of the local gradient ${\bf{s}}_{t,k}$ and the signal vector of the data samples ${\bf{d}}_{t,k}=[d_{t,k,1},\ldots,d_{t,k,Q}]^{\rm T} \in \mathbb{R}^Q$ should be aligned to have the same number of symbols, as they share the same time-frequency resources.
As presented before, each dimension of the local gradient is normalized to a symbol of ${\bf{s}}_{t,k}$.
To align ${\bf{d}}_{t,k}$ with ${\bf{s}}_{t,k}$, devices appropriately map multiple bits of the data samples to a symbol of ${\bf{d}}_{t,k}$, and then apply a proper zero padding scheme~\cite{Mohammadkarimi2018Signature}.
The $q$-th entry of ${\bf{d}}_{t,k}$ yields $\mathbb{E}[d_{t,k,q}]=0$ and $\mathbb{E}[d_{t,k,q}^2]=1$.
For simplicity, we assume that the entries of ${\bf{s}}_{t,k}$ and ${\bf{d}}_{t,k}$ are independent of each other~\cite{Qi2021Integrated}, i.e., $\mathbb{E}[s_{t,k,q}d_{t,k,q}]=0,q=1,2,\ldots,Q,\forall k \in \mathcal{K}$.
Each communication round is equally divided into $Q$ slots.
In the $q$-th slot of the $t$-th communication round, the devices transmit the superposition of $\{s_{t,k,q}\}$ and $\{d_{t,k,q}\}$ to the BS after being processed by the parallel-to-serial (P/S) conversion.

At the BS side, the superposition signal received by each antenna in the $q$-th slot of the $t$-th communication round is independently downconverted to form the baseband superposition signal vector ${\bf{y}}_{t,q}$, as given by
\begin{align}
	{\bf{y}}_{t,q} =\! \underbrace{\sum \nolimits_{k=1}^K { {p_{t,f,k}} {{\bf{h}}_{t,k}} s_{t,k,q} }}_{\text{local gradients}} +\! \underbrace{\sum \nolimits_{k=1}^K { {p_{t,c,k}} {{\bf{h}}_{t,k}} d_{t,k,q} }}_{\text{data samples}} +\! \underbrace{{\bf{n}}_{t,q}}_{\text{noise}},
\end{align}
where $p_{t,f,k} \in \mathbb{C}$ and $p_{t,c,k} \in \mathbb{C}$ are the transmit power allocation coefficients for local gradients and data samples, respectively, ${\bf{n}}_{t,q} \sim \mathcal{CN}({\bf{0}}, {{{\sigma}^2}}{{\bf{I}}_{N_r}} )$ is the additive white Gaussian noise, and ${\bf{h}}_{t,k} \in \mathbb{C}^{{N_r}}$ is the channel coefficient vector from the $k$-th device to the BS.
We consider a block-fading channel, where ${\bf{h}}_{t,k}$ remains unchanged within a communication round but varies independently between rounds, and assume the perfect channel state information is available~\cite{Cao2022Transmission}.

As shown in Fig.~\ref{block_diagram}, to decode data samples, the BS performs receive beamforming~\cite{Dutta2020ACase} by multiplying ${\bf{y}}_{t,q}$ with a beamforming matrix containing $K$ beamformers, i.e., ${\bf{F}}_t\triangleq [{\bf{f}}_{t,1},{\bf{f}}_{t,2},\ldots,{\bf{f}}_{t,K}] \in \mathbb{C}^{N_r \times K}$, thereby generating $K$ separated parallel data streams denoted by a vector $\hat{\bf{d}}_{t,q}={\bf{F}}_t^{\rm H}{\bf{y}}_{t,q}={[\hat{d}_{t,1,q},\hat{d}_{t,2,q},\ldots,\hat{d}_{t,K,q}]}^{\rm T} \in \mathbb{C}^{K}$~\cite{Qi2021Integrated,Zhu2022Enhanced}.
The $k$-th data stream for decoding the data samples from the $k$-th device is given by
\begin{align}
		\hat{d}_{t,k,q} \!=&{\bf{f}}^{\rm H}_{t,k}{\bf{y}}_{t,q}\notag \\
		=&{\bf{f}}_{t,k}^{\rm H}({p_{t,c,k}}{\bf{h}}_{t,k}d_{t,k,q})\!\! +\! \underbrace{{\bf{f}}_{t,k}^{\rm H} \left(\sum \nolimits_{k'=1}^K \! {{p_{t,f,k'}}{{\bf{h}}_{t,k'}} s_{t,{k'},q}}\right)}_{\text{interference of local gradients}} \notag \\ 
		&+ \underbrace{{\bf{f}}_{t,k}^{\rm H} \left(\sum\nolimits_{{k'}=1,k' \neq k}^K \! {p_{t,c,k'}{\bf{h}}_{t,k'}d_{t,{k'},q}}\right)}_{\text{interference of other data samples}} \notag \\ 
		&+ {\bf{f}}_{t,k}^{\rm H}{\bf{n}}_{t,q}, \forall k \in \mathcal{K},
\end{align}
\noindent which is independent of other data streams for decoding.
As a result, the signal-to-interference-plus-noise ratio (SINR) of the $k$-th device, $\gamma_{t,k}$, is presented by (\ref{SINR}) at the top of this page.
\begin{figure*}[t]
	\begin{align}
		\label{SINR}
		\gamma_{t,k}=\frac{|{\bf{f}}_{t,k}^{\rm H}({p_{t,c,k}}{\bf{h}}_{t,k})|^2}{\sum\nolimits_{{k'}=1,{k' \neq k}}^K {|{\bf{f}}_{t,k}^{\rm H}(p_{t,c,k'}{\bf{h}}_{t,k'})|^2} + \sum \nolimits_{k'=1}^K { \frac{N_{f,k'}^2}{N_f^2} |{\bf{f}}_{t,k}^{\rm H}({p_{t,f,k'}}{{\bf{h}}_{t,k'}})|^2} + {\sigma^2}\|{\bf{f}}_{t,k}\|^2 },~\forall k \in \mathcal{K}.
	\end{align}
	\hrulefill
\end{figure*}
The decoded symbols are accumulated over $Q$ slots to recover signal vectors $\{{\bf{d}}_{t,k}\}$ using the serial-parallel (S/P) converter, and the uploaded data samples are recovered by post-processing.

After removing all data sample signals from the superposition signal vector ${\bf{y}}_{t,q}$, the residual local gradients are free from the interference of data samples.
Then, the BS employs another beamformer ${\bf{b}}_t\in{\mathbb{C}^{N_r}}$ to aggregate the local gradients over the air, which is given by
\begin{align}
	\label{aggregated_gradient}
	\hspace{-0.1 cm}{\hat{s}_{t,q}}\!=\!  \sum \nolimits_{k=1}^K { {p_{t,f,k}} {\bf{b}}_t^{\rm H} {{\bf{h}}_{t,k}} s_{t,k,q} } + {\bf{b}}_t^{\rm H}{\bf{n}}_{t,q}.
\end{align}
Since the desired aggregation signal is $s_{t,q}=\sum\nolimits_{k=1}^K {s_{t,k,q}}$, the distortion between $s_{t,q}$ and $\hat{s}_{t,q}$ is measured by the MSE:
\begin{align}
	\label{MSE}
	{\rm MSE}_t =&\mathbb{E}[|\hat{s}_{t,q}-s_{t,q}|^2] \notag \\
	=& \sum\nolimits_{k=1}^K {\frac{N_{f,k}^2}{N_f^2}\left|p_{t,f,k}{\bf{b}}_t^{\rm H}{\bf{h}}_{t,k}-1\right|^2} + \|{\bf{b}}_t\|^2\sigma^2.
\end{align}
Similarly, the aggregated gradient signals accumulated over $Q$ slots are rearranged in an estimation signal vector $\hat{\bf{s}}_t=[\hat{s}_{t,1},\ldots,\hat{s}_{t,Q}]^{\rm T} \in \mathbb{C}^Q$ using the S/P converter.
Finally, the BS post-processes $\hat{\bf{s}}_t$ to obtain the aggregated gradient by de-normalization, ${\hat{g}^f_{t,q}}={\bar{\sigma}_t}{\hat{s}_{t,q}} + {\bar{g}_t}$, i.e.,
\begin{align}
	\label{actual_aggregated_model}
	\hspace{-0.3 cm}{\hat{g}^f_{t,q}}=& \underbrace{ \sum \nolimits_{k=1}^K { {\frac{N_{f,k}}{N_f}}\left(1-{p_{t,f,k}} {\bf{b}}_t^{\rm H} {{\bf{h}}_{t,k}}\right){\bar{g}_t} }\! +\! {\bar{\sigma}_t}{\bf{b}}_t^{\rm H}{\bf{n}}_{t,q}}_{\text{de-normalization error due to the channel fading and noise}} \notag \\
	&+ \sum \nolimits_{k=1}^K \! {{\frac{N_{f,k}}{N_f}}{p_{t,f,k}} {\bf{b}}_t^{\rm H} {{\bf{h}}_{t,k}}g^f_{t,k,q}}, q=1,2,\ldots,Q,
\end{align}
\noindent where ${\hat{g}^f_{t,q}}$ is the $q$-th entry of the aggregated gradient of FL, i.e., ${\hat{\bf{g}}_t^f}=[\hat{g}^f_{t,1},\ldots,\hat{g}^f_{t,Q}]^{\rm T}$.
	
\section{Convergence Analysis and Problem Formulation}
\label{convergence_analysis_and_problem_formulation}
In this section, we provide the convergence analysis of the proposed SemiFL by characterizing the optimality gap based on commonly adopted assumptions.
Then, we formulate a problem to minimize the optimality gap by jointly optimizing the transmit power allocation coefficients $\{p_{t,f,k}\}$ and $\{p_{t,c,k}\}$, the aggregation beamformer ${\bf{b}}_t$, and the decoding beamformers $\{{\bf{f}}_{t,k}\}$.

\subsection{Convergence Analysis}
\label{convergence_analysis}
To facilitate the convergence analysis, we impose the following standard assumptions on the global empirical loss function $F({\bf{w}})$ and gradients, which have been extensively employed by the works in~\cite{Chen2021A,Liu2021Reconfigurable,Amiri2021Blind,Fan2021Joint,Cao2022Transmission}.

\begin{assumption}[$\mu$-strongly convex]
	\label{assumption_1_text}
	The global empirical loss function $F(\bf{w})$ is $\mu$-strongly convex.
	Therefore, for any ${\bf{w}}$, ${\bf{w}}' \in \mathbb{R}^{Q}$ and $\mu > 0 $, we have
	\begin{align}
		\label{assumption_1}
		\hspace{-0.15 cm}F({\bf{w}}) \ge F({\bf{w}}') + \left({\bf{w}} - {\bf{w}}' \right)^{\rm T}{\nabla F({\bf{w}}')} + \frac{\mu}{2} \|{\bf{w}} - {\bf{w}}'\|^2,
	\end{align}
	where $\nabla F({\bf{w}})$ is the gradient of the global empirical loss function $F({\bf{w}})$ regarding $\bf{w}$.
\end{assumption}

\begin{assumption}[$L$-smooth]
	\label{assumption_2_text}
	The global empirical loss function $F(\bf{w})$ is $L$-smooth.
	Therefore, for any ${\bf{w}}$, ${\bf{w}}'  \in \mathbb{R}^{Q}$ and $L > 0$, we have
	\begin{align}
		\label{assumption_2}
		\hspace{-0.2 cm}F({\bf{w}}) \le F({\bf{w}}') + \left({\bf{w}} - {\bf{w}}' \right)^{\rm T}{\nabla F({\bf{w}}')} + \frac{L}{2} \|{\bf{w}} - {\bf{w}}'\|^2.
	\end{align}
\end{assumption}

\begin{assumption}[Bounded gradients]
	\label{assumption_3_text}
	The squared $2$-norms of any local gradient and any sample-wise gradient are bounded. 
	Therefore, for constants $G^2\ge0$, $\xi_1\ge0$ and $\xi_2>0$, we have
	\begin{align}
		\label{assumption_3}
		&\mathbb{E}{\left[ \|{\bf{g}}^f_{t,k}\|^2 \right]} \! \le \! G^2,~\forall k \in \mathcal{K},,~\forall t, \\
		&\|{\bf{g}}_{t,k,n}\|^2\! \le \! \xi_1 \! + \! \xi_2 \|\nabla F({\bf{w}}_t)\|^2,~\forall k \in \mathcal{K},~\forall n \in \mathcal{D},~\forall t.
	\end{align}
\end{assumption}
\noindent Assumption~\ref{assumption_1_text} can be the foundation for deriving the celebrated Polyak-Lojasiewicz (PL) inequality~\cite{Ni2022STAR}, which will be utilized in Appendix~\ref{proof_of_theorem_1}.
Assumption~\ref{assumption_3_text} bounds the norms of gradients to facilitate the scaling operations during derivation.

The convergence analysis starts with characterizing the error of the global gradient $\hat{\bf{g}}_{t}$ in the $t$-th round.
By rewriting $\hat{\bf{g}}_{t}$ as $\hat{\bf{g}}_{t}={\nabla F({\bf{w}}_t)} - {\bf{e}}$ and plugging it into the gradient descent method mentioned at the end of Section~\ref{federated_leanring_model}, the global model update is rewritten as
\begin{align}
	\label{actual_model_update}
	{\bf{w}}_{t+1} = {\bf{w}}_t - \eta \left( {\nabla F({\bf{w}}_t)} - {\bf{e}} \right),
\end{align}
where $\bf{e}$ denotes the error of the global gradient, given by
\begin{align}
	\label{definition_of_e}
	{\bf{e}} \!=\! a_1( \underbrace{{\nabla F({\bf{w}}_t)}\!\!-\!{\bf{g}}_t^f}_{{\bf{e}}_1} )\! +\! a_2 (\underbrace{{\nabla F({\bf{w}}_t)} \!\!-\! {\bf{g}}_t^c}_{{\bf{e}}_2} ) \!+\! a_1 ( \underbrace{{\bf{g}}_t^f \!\!-\! {\hat{\bf{g}}}_t^f}_{{\bf{e}}_3} ).
\end{align}
Here, $a_1=\frac{N_f}{N_f+N_c}$, $a_2=\frac{N_c}{N_f+N_c}$, ${\bf{g}}_t^f=\sum\nolimits_{k=1}^{K}{\frac{N_{f,k}}{N_f}{\bf{g}}^f_{t,k}}$ is the desired FL gradient without considering the aggregation error, and ${\nabla F({\bf{w}}_t)}=\frac{1}{N} \sum\nolimits_{k=1}^K \sum_{n \in \mathcal{D}_k} {{\bf{g}}_{t,k,n}}$ is the ideal global gradient of the global empirical loss function.
One can find that the global gradient error $\bf{e}$ can be decomposed into three parts, i.e., ${\bf{e}}_1$, ${\bf{e}}_2$, and ${\bf{e}}_3$, which separately capture the primary hostilities jeopardizing the convergence behavior of SemiFL.
Specifically, the error ${\bf{e}}_1$ is due to the limited computing capabilities of devices, which restricts the amount of data samples for calculating local gradients.
Randomly selecting a limited number of samples from all accumulated data samples at the BS compromises the CL gradient and thus incurs the error ${\bf{e}}_2$.
The error ${\bf{e}}_3$ captures the impact of the undesirable channel fading and noise on the aggregated gradient.

Based on~\cite{Liu2021Reconfigurable} and~\cite{Friedlander2012Hybrid}, we now reveal how the error $\bf{e}$ affects the convergence behavior of SemiFL between two consecutive rounds in the following lemma.
\begin{lemma}
	\label{lemma_1}
	Suppose $F({\bf{w}}_t)$ satisfies Assumption 2, and let learning rate $\eta=1/L$. In the $t$-th round, we have
	%
	\begin{align}
		\label{consecutive_convergence}
		\mathbb{E}\left[ F({\bf{w}}_{t+1}) \right] \le& \mathbb{E}\left[ F({\bf{w}}_t) \right] - \frac{1}{2L}\|{\nabla F({\bf{w}}_t)}\|^2 + \frac{2{a_1^2}}{L}\mathbb{E}[ \|{\bf{e}}_1 \|^2 ] \notag\\
		&+ \frac{2a_2^2}{L}\mathbb{E}[ \|{\bf{e}}_2 \|^2 ] + \frac{a_1^2}{L}\mathbb{E}[ \|{\bf{e}}_3 \|^2 ].
	\end{align}
\end{lemma}
\begin{IEEEproof}
	Please refer to Appendix~\ref{proof_of_lemma_1}.
\end{IEEEproof}

Next, we bound $\mathbb{E}[ \|{\bf{e}}_1\|^2 ]$, $\mathbb{E}[ \|{\bf{e}}_2\|^2 ]$ and $\mathbb{E}[ \|{\bf{e}}_3\|^2 ]$ based on Assumption~\ref{assumption_3_text}, as presented in the following lemma.
\begin{lemma}
	\label{lemma_2}
	Given Assumption~\ref{assumption_3_text}, the squared $2$-norms of the errors, i.e., $\mathbb{E}[ \|{\bf{e}}_1\|^2 ]$, $\mathbb{E}[ \|{\bf{e}}_2\|^2 ]$, and $\mathbb{E}[ \|{\bf{e}}_3\|^2 ]$, are upper bounded, respectively, by
	\begin{align}
		\label{bound_of_e1}
		\mathbb{E}\left[ \|{\bf{e}}_1\|^2 \right] \le& \frac{N-N_f}{N_f}(\xi_1 + \xi_2\|{\nabla F({\bf{w}}_t)}\|^2), \\
		\label{bound_of_e2}
		\mathbb{E}\left[ \|{\bf{e}}_2\|^2 \right] \le& \frac{N-N_c}{N_c}(\xi_1 + \xi_2\|{\nabla F({\bf{w}}_t)}\|^2), \\
		\label{bound_of_e3}
		\mathbb{E}\left[ \|{\bf{e}}_3\|^2 \right] \le& \frac{4KG^2}{N_f^2} \sum\nolimits_{k=1}^K  {N_{f,k}^2 \left| 1 - p_{t,f,k} {\bf{b}}_t^{\rm H} {\bf{h}}_{t,k} \right|^2}\notag \\ 
		&+ G^2 \sigma^2 \|{\bf{b}}_t\|^2.
	\end{align}
\end{lemma}
\begin{IEEEproof}
	Please refer to Appendix~\ref{proof_of_lemma_2}.
\end{IEEEproof}

Finally, based on Lemmas~\ref{lemma_1} and~\ref{lemma_2}, we characterize the convergence behavior of SemiFL framework by deriving the optimality gap in Theorem~\ref{theorem_1}.
\begin{theorem}[Optimality gap of SemiFL]
	\label{theorem_1}
	Suppose Assumptions~\ref{assumption_1_text},~\ref{assumption_2_text}, and~\ref{assumption_3_text} hold and set learning rate $\eta=\frac{1}{L}$. 
	Let ${\bf{w}}^*$ denote the optimal global model.
	Then, the optimality gap of SemiFL after $T$ rounds is given by:
	%
	\begin{align}
		\label{convergence_bound}
		&\mathbb{E}[ F({\bf{w}}_{T+1}) \!-\!\! F({\bf{w}}^*) ] \le \rho_1^T \mathbb{E}\left[ F({\bf{w}}_1) - F({\bf{w}}^*) \right] \!+\! \rho_2 \frac{1\!-\!\rho_1^T}{1\!-\!\rho_1} \notag\\
		&+ \sum \nolimits_{t=1}^{T}  {\rho_1^{T-t} \varphi_{t}\left( \{ p_{f,k} \}, {\bf{b}} \right)}  \triangleq  \psi_T^{\rm SemiFL}\! \left( \{ p_{f,k} \}, {\bf{b}} \right),
	\end{align}
	where $\rho_1 = 1-\frac{\mu}{L}+ 4 \mu \xi_2 \frac{N_f (N-N_f) + N_c (N-N_c)}{L(N_f+N_c)^2}$, $\rho_2 = 2\xi_1 \frac{N_f(N-N_f)+N_c(N-N_c)}{L(N_f+N_c)^2}$, and $\varphi_{t}\left( \{ p_{f,k} \}, {\bf{b}} \right)\! =  \frac{4KG^2 \sum\nolimits_{k=1}^K { N_{f,k}^2 \left| 1 - p_{t,f,k} {\bf{b}}_t^{\rm H} {\bf{h}}_{t,k} \right|^2}}{L(N_f+N_c)^2} +\frac{N_f^2G^2\sigma^2\|{\bf{b}}_t\|^2}{L(N_f+N_c)^2}$.
\end{theorem}
\begin{IEEEproof}
	Please refer to Appendix~\ref{proof_of_theorem_1}.
\end{IEEEproof}

\begin{remark}[The value range of $\xi_2$]
	\label{remark_1}
	Note $\xi_2$ should be in the range $(0, \frac{(N_f+N_c)^2}{4[N_f(N-N_f)+N_c(N-N_c)]})$ to guarantee the convergence of $\mathbb{E}[ F({\bf{w}}_{T+1}) - F({\bf{w}}^*) ]$, while ensuring the correctness of applying the PL inequality in (\ref{consecutive_convergence_1}).
	The reasons are three-fold:
	\begin{enumerate}
		\item Since the stable convergence of $\psi_T^{\rm SemiFL}\left( \{ p_{f,k} \}, {\bf{b}} \right)$ requires for $0<\rho_1<1$, we have
		\begin{align}
			(1-\frac{L}{\mu})\frac{(N_f+N_c)^2}{4[N_f(N-N_f)+N_c(N-N_c)]}<\xi_2\notag \\
			<\frac{(N_f+N_c)^2}{4[N_f(N-N_f)+N_c(N-N_c)]}.
		\end{align}
		\item The term $\frac{1}{2L} - 2\xi_2\frac{N_f(N-N_f)+N_c(N-N_c)}{L(N_f+N_c)^2}$ in (\ref{consecutive_convergence_1}) should be non-negative, which implies
		\begin{align}
			\xi_2\le\frac{(N_f+N_c)^2}{4[N_f(N-N_f)+N_c(N-N_c)]}.
		\end{align}
		\item It is known from Assumption~\ref{assumption_3_text} that $\xi_2>0$.
	\end{enumerate}
	As a result, we have $0<\xi_2 < \frac{(N_f+N_c)^2}{4[N_f(N-N_f)+N_c(N-N_c)]}$.
	For a constant $\xi_2$ greater than the threshold $\frac{(N_f+N_c)^2}{4[N_f(N-N_f)+N_c(N-N_c)]}$, one can enlarge the threshold by increasing $N_c$ to satisfy the condition.
\end{remark}

We extend the result in Theorem~1 to a special case where the BS calculates the CL gradient using all data samples accumulated in previous rounds, as given in Corollary~\ref{corollary_new_1}.
\begin{corollary}[Optimality gap using all accumulated data]
	\label{corollary_new_1}
	Given Assumptions~\ref{assumption_1_text},~\ref{assumption_2_text}, and~\ref{assumption_3_text} as well as the learning rate $\eta=\frac{1}{L}$, suppose the BS uses all accumulated $\bar{N}_{c,t}=tN_c$ data samples till the $t$-th round to calculate the CL gradient. Then, the optimality gap of SemiFL after $T$ rounds is given by:
	\begin{align}
		\label{optimali_gap_all_samples}
		&\mathbb{E}[ F({\bf{w}}_{T+1}) - F({\bf{w}}^*) ] \le {\left(\prod\limits_{t=1}^{T} \bar{\rho}_{1,t}\right)} \mathbb{E}\left[ F({\bf{w}}_1) - F({\bf{w}}^*) \right] \notag \\ 
		&+ \sum\limits_{t=1}^{T} \left( \prod\limits_{i=t+1}^{T} \bar{\rho}_{1,i} \right) \bar{\rho}_{2,t} +\sum \limits_{t=1}^{T} { \left( \prod\limits_{i=t+1}^{T}\! \bar{\rho}_{1,i} \right) \bar{\varphi}_{t}\left( \{ p_{f,k} \}, {\bf{b}} \right)}\notag \\ 
		&\triangleq \bar{\psi}_{T}^{\rm SemiFL}\left( \{ p_{f,k} \}, {\bf{b}} \right),
	\end{align}
	where $\bar{\rho}_{1,t}=1-\frac{\mu}{L}+4\mu\xi_2 \frac{N_f (N-N_f) + \bar{N}_{c,t} (N-\bar{N}_{c,t})}{L(N_f+\bar{N}_{c,t})^2}$, $\bar{\rho}_{2,t}=2\xi_1 \frac{N_f (N-N_f) + \bar{N}_{c,t} (N-\bar{N}_{c,t})}{L(N_f+\bar{N}_{c,t})^2}$, and $\bar{\varphi}_{t} \left( \{ p_{f,k} \}, {\bf{b}} \right)=\frac{4KG^2 \sum\nolimits_{k=1}^K { N_{f,k}^2 \left| 1 - p_{t,f,k} {\bf{b}}_t^{\rm H} {\bf{h}}_{t,k} \right|^2}}{L(N_f+\bar{N}_{c,t})^2} +\frac{N_f^2G^2\sigma^2\|{\bf{b}}_t\|^2}{L(N_f+\bar{N}_{c,t})^2}$.
\end{corollary}
\begin{IEEEproof}
	Please refer to Appendix~\ref{proof_corollary_new_1}.
\end{IEEEproof}
Since $\frac{N_f (N-N_f) + \bar{N}_{c,t} (N-\bar{N}_{c,t})}{L(N_f+\bar{N}_{c,t})^2} \le \frac{N_f (N-N_f) + N_c (N-N_c)}{L(N_f+N_c)^2}$ and $(N_f+\bar{N}_{c,t})^2 \ge (N_f+N_c)^2, \forall t \ge 1$, it can be verified that $\rho_1 \ge \bar{\rho}_{1,t}$, $\rho_2 \ge \bar{\rho}_{2,t}$, and $\varphi_{t}\left( \{ p_{f,k} \}, {\bf{b}} \right) \ge \bar{\varphi}_{t}\left( \{ p_{f,k} \}, {\bf{b}} \right), \forall t \ge 1$. Hence, we have $\bar{\psi}_{T}^{\rm SemiFL}\left( \{ p_{f,k} \}, {\bf{b}} \right) \le \psi_{T}^{\rm SemiFL}\left( \{ p_{f,k} \}, {\bf{b}} \right)$. Corollary~\ref{corollary_new_1} indicates that using all accumulated samples for calculating CL gradient contributes to a smaller optimality gap of SemiFL.
Furthermore, we extend the optimality gap of SemiFL in Theorem~\ref{theorem_1} to another specific case where a decreasing learning rate is adopted~\cite{Guo2021Analog}, as given in Corollary~\ref{corollary_new_2}.
\begin{corollary}[Optimality gap with a decreasing learning rate]
	\label{corollary_new_2}
	Given Assumptions~\ref{assumption_1_text} and~\ref{assumption_2_text}, suppose the decreasing learning rate in the $t$-th round is designed as $\eta_t=\frac{\Lambda}{t+\tau}$, where $\Lambda > \frac{1}{\mu}$, and $\tau \ge \Lambda L$.
	Then, the optimality gap of SemiFL after $T$ rounds is given by:
	\begin{align}
		\label{optimality_gap_decreasing_learning_rate}
		\mathbb{E}[F({\bf{w}}_{T+1})-F({\bf{w}}^*)] \le \frac{\theta_T}{T+\tau+1},
	\end{align}
	where $\theta_T=\max\{\frac{\Lambda (T+\tau) \mathbb{E}[\|{\bf{e}}_T\|^2]}{2 (\Lambda \mu - 1)},\mathbb{E}[F({\bf{w}}_T)-F({\bf{w}}^*)](T+\tau)\}$ and ${\bf{e}}_T = \nabla F({\bf{w}}_T) - \hat{\bf{g}}_T$.
	When $T \rightarrow \infty$, one can obtain $\mathbb{E}[F({\bf{w}}_{T+1})-F({\bf{w}}^*)] \rightarrow 0$.
\end{corollary}
\begin{IEEEproof}
	Please refer to Appendix~\ref{proof_corollary_new_2}.
\end{IEEEproof}

From Corollary~\ref{corollary_new_2}, it can be confirmed that the optimality gap of SemiFL converges to $0$ with the convergence rate $\mathcal{O}(\frac{1}{T})$ when applying a decreasing learning rate for a sufficiently large $T$.
Based on Theorem~\ref{theorem_1}, Corollary~\ref{corollary_new_1}, and Corollary~\ref{corollary_new_2}, the convergence of the proposed SemiFL under different settings is proved.

In order to provide thorough insights into the relation between SemiFL, FL, and CL, we first capture the convergence behavior of FL and CL, and then compare them with SemiFL in Theorem~\ref{theorem_2}.
For comparison fairness, we stipulate that all devices utilize $N_f+N_c$ data samples to calculate the local gradient in FL but transmits no data to the BS.
Consequently, the global model is trained by the aggregated gradient only.
For CL, we consider that all devices only upload $N_f+N_c$ data samples to the BS in each round but never perform local training.
Accordingly, the BS randomly selects $N_f+N_c$ data samples from its accumulated data to calculate the CL gradient to update the global model.
\begin{theorem}[Relation between SemiFL, FL, and CL]
	\label{theorem_2}
	Given Assumptions~\ref{assumption_1_text},~\ref{assumption_2_text}, and~\ref{assumption_3_text} as well as the learning rate $\eta=\frac{1}{L}$, the optimality gaps of FL and CL after $T$ rounds are given by \emph{(\ref{convergence_bound_FL})} and \emph{(\ref{convergence_bound_CL})}, respectively.
	\begin{align}
		\label{convergence_bound_FL}
		&\mathbb{E}[ F({\bf{w}}_{T+1}) \!-\! F({\bf{w}}^*) ] \le \tilde{\rho}_1^T \mathbb{E}\left[ F({\bf{w}}_1) \!-\! F({\bf{w}}^*) \right] \!+\! \tilde{\rho}_2 \frac{1-\tilde{\rho}_1^T}{1-\tilde{\rho}_1} \notag \\
		&+\sum \nolimits_{t=1}^{T} {\tilde{\rho}_1^{T-t} \tilde{\varphi}_{t}\left( \{ p_{f,k} \}, {\bf{b}} \right)} \triangleq \psi_T^{\rm FL}\left( \{ p_{f,k} \}, {\bf{b}} \right), \\
		\label{convergence_bound_CL}
		&\mathbb{E}[ F({\bf{w}}_{T+1})\! -\! F({\bf{w}}^*) ] \le \hat{\rho}_1^T \mathbb{E}\left[ F({\bf{w}}_1)\! -\! F({\bf{w}}^*) \right]\! +\! \hat{\rho}_2 \frac{1-\hat{\rho}_1^T}{1-\hat{\rho}_1} \notag \\ 
		&\quad\quad\quad\quad\quad\quad\quad\quad\quad\triangleq \psi_T^{\rm CL},
	\end{align}
	where $\tilde{\rho}_1 = 1-\frac{\mu}{L}+ 8 \mu \xi_2 \frac{N-(N_f+N_c)}{L(N_f+N_c)}$, $\tilde{\rho}_2 = 4\xi_1 \frac{N-(N_f+N_c)}{L(N_f+N_c)}$, $\hat{\rho}_1 = 1-\frac{\mu}{L}+ \mu \xi_2 \frac{N-(N_f+N_c)}{L(N_f+N_c)}$, $\hat{\rho}_2 = \xi_1 \frac{N-(N_f+N_c)}{2L(N_f+N_c)}$, and $\tilde{\varphi}_{t}\left( \{ p_{f,k} \}, {\bf{b}} \right) \! = \frac{G^2\sigma^2\|{\bf{b}}_t\|^2}{L} + \frac{4KG^2 \sum\nolimits_{k=1}^K { (N_{f,k}+N_{c,k})^2 \left| 1 - p_{t,f,k} {\bf{b}}_t^{\rm H} {\bf{h}}_{t,k} \right|^2}}{L(N_f+N_c)^2}$.
	Then, we have the following relation between SemiFL, FL, and CL:
	\begin{align}
		\label{convergence_comparison}
		\hspace{-0.1 cm}\psi_T^{\rm CL}\! \le \! \psi_T^{\rm SemiFL}\!\left( \{ p_{f,k} \}, {\bf{b}} \right)\! \le \! \psi_T^{\rm FL}\!\left( \{ p_{f,k} \}, {\bf{b}} \right).\!
	\end{align}
\end{theorem}
\begin{IEEEproof}
	Please refer to Appendix~\ref{proof_of_theorem_2}.
\end{IEEEproof}

On the one hand, thanks to the CL empowered by the computing capability of the BS, Theorem~\ref{theorem_2} proves that SemiFL outperforms FL by achieving a smaller optimality gap.
On the other hand, CL achieves the smallest optimality gap among the three learning frameworks, which can be regarded as a performance upper bound.
When retaining all $N_c+N_f$ data samples for local training, we obtain the optimality gap of SemiFL, i.e., $\psi_T^{\rm SemiFL}( \{ p_{f,k} \}, {\bf{b}} )$, degenerates into that of FL, i.e., $\psi_T^{\rm FL}( \{ p_{f,k} \}, {\bf{b}} )$.
When dedicating all $N_c+N_f$ data samples to CL and ignoring the impact of wireless channels on gradient aggregation, the optimality gap of SemiFL, i.e., $\psi_T^{\rm SemiFL}( \{ p_{f,k} \}, {\bf{b}} )$, reduces to that of CL, i.e., $\psi_T^{\rm CL}$.
Therefore, Theorem~\ref{theorem_2} theoretically confirms that the proposed SemiFL is a more general learning paradigm than FL and CL.
\begin{remark}[Impact of wireless communication]
	\label{remark_2}
	Based on Theorems~\ref{theorem_1} and~\ref{theorem_2}, we have the following observations about the impact of wireless communication on the optimality gaps.
	\begin{itemize}
		\item As T goes to infinity, the optimality gaps of SemiFL, FL, and CL tend to the following three limits, respectively:
		\vspace{-0.2 cm}
		\begin{align}
			\label{limits_SemiFL}
			&\lim_{T \rightarrow \infty} {\psi_T^{\rm SemiFL} \!\! \left( \{ p_{f,k} \}, {\bf{b}} \right)} = \lim_{T \rightarrow \infty} {\sum \limits_{t=1}^{T} {\rho_1^{T-t} \varphi_{t} \! \left( \{ p_{f,k} \}, {\bf{b}} \right)}} \notag \\
			&\hspace{+4 cm}+ \frac{\rho_2}{1-\rho_1}, \\
			\label{limits_FL}
			&\lim_{T \rightarrow \infty} {\psi_T^{\rm FL} \! \left( \{ p_{f,k} \}, {\bf{b}} \right)} =  \lim_{T \rightarrow \infty} {\sum \limits_{t=1}^{T} {\tilde{\rho}_1^{T-t} \tilde{\varphi}_{t} \! \left( \{ p_{f,k} \}, {\bf{b}} \right)}} \notag \\
			&\hspace{+3.4 cm}+ \frac{\tilde{\rho}_2}{1-\tilde{\rho}_1}, \\
			\label{limits_CL}
			&\lim_{T \rightarrow \infty}\! {\psi_T^{\rm CL}} = \frac{\hat{\rho}_2}{1-\hat{\rho}_1}.
		\end{align}
		Due to the detrimental impact of the undesirable wireless communication contained in $\varphi_t\left( \{ p_{f,k} \}, {\bf{b}} \right)$ and $\tilde{\varphi}_t\left( \{ p_{f,k} \}, {\bf{b}} \right)$, both $\psi_T^{\rm SemiFL} \! \left( \{ p_{f,k} \}, {\bf{b}} \right)$ and $\psi_T^{\rm FL} \! \left( \{ p_{f,k} \}, {\bf{b}} \right)$ are fluctuating and can not converge to a stable value even if $T$ goes to infinity.
		This reveals the significance and necessity of designing the transceivers, i.e., the transmit power allocation and receive beanformers, to reduce the optimality gap.
		\item The wireless factors in the distant past have marginal impacts on the optimality gap than the recent ones.
		Since $0<\rho_1<1$ and $0<\tilde{\rho}_1<1$, it is obtained that $\tilde{\varphi}_t\left( \{ p_{f,k} \}, {\bf{b}} \right)$ and $\varphi_t\left( \{ p_{f,k} \}, {\bf{b}} \right)$ in early rounds have much smaller weight coefficients based on \emph{(\ref{convergence_bound})} and \emph{(\ref{convergence_bound_FL})}.
		This coincides with the observations in~\emph{\cite{Cao2022Transmission}}.
	\end{itemize}
\end{remark}

Furthermore, we also derive the optimality gaps of SemiFL and FL over error-free wireless channels in the following corollary.
Note that the error-free wireless channels refer to the case where there is no communication noise and the wireless-related factors are perfectly designed such that local gradients are aggregated without any error.

\begin{corollary}[Optimality gaps over error-free channels]
	\label{corollary_1}
	Given Assumptions~\ref{assumption_1_text},~\ref{assumption_2_text}, and~\ref{assumption_3_text} as well as learning rate $\eta=\frac{1}{L}$, the optimality gap of SemiFL and FL over error-free wireless channels are given, respectively, by
	\vspace{-0.2 cm}
	\begin{align}
		\label{error-free_convergence_bound}
		\mathbb{E}\left[ F({\bf{w}}_{T+1}) \!-\! F({\bf{w}}^*) \right] \le \rho_1^T \mathbb{E}\left[ F({\bf{w}}_1)\! -\! F({\bf{w}}^*) \right]\! +\! \rho_2 \frac{1-\rho_1^T}{1-\rho_1}, \\
		\label{error-free_convergence_bound_FL}
		\mathbb{E}\left[ F({\bf{w}}_{T+1}) \!-\! F({\bf{w}}^*) \right] \le \tilde{\rho}_1^T \mathbb{E}\left[ F({\bf{w}}_1)\! -\! F({\bf{w}}^*) \right]\! +\! \tilde{\rho}_2 \frac{1-\tilde{\rho}_1^T}{1-\tilde{\rho}_1},
	\end{align}
\end{corollary}
\begin{IEEEproof}
	In light of the description of error-free wireless channels, we have $\|{\bf{e}}_3\|^2 = 0$.
	By plugging (\ref{bound_of_e1}), (\ref{bound_of_e2}) and $\|{\bf{e}}_3\|^2 = 0$ into (\ref{consecutive_convergence}), while recursively applying the result for $T$ times, we reach (\ref{error-free_convergence_bound}).
	Similarly, the optimality gap (\ref{error-free_convergence_bound_FL}) can be obtained by plugging (\ref{FL_gradient_bound_1}) and $\|\tilde{\bf{e}}_3\|^2 = 0$ into (\ref{FL_consecutive_fundation_1}) and recursively applying the result for $T$ times.
\end{IEEEproof}

Based on Corollary~\ref{corollary_1} and (\ref{limits_CL}), we observe that, in the error-free case, all three schemes of SemiFL, FL, and CL can converge to the optimal model without any gaps under certain conditions.
Specifically, as $T\rightarrow \infty$, we have the optimality gaps without aggregation errors of SemiFL, FL, and CL tend to be $\frac{\rho_2}{1-\rho_1}$, $\frac{\tilde{\rho}_2}{1-\tilde{\rho}_1}$, and $\frac{\hat{\rho}_2}{1-\hat{\rho}_1}$, respectively.
For SemiFL, if the relation between $N_f$ and $N_c$ satisfies ${(N_f\!-\!\frac{N}{2})}^2\!+\!{(N_c\!-\!\frac{N}{2})}^2\!=\!\frac{N^2}{2}$, we observe $\frac{\rho_2}{1-\rho_1} \! \rightarrow \! 0$, i.e., the optimality gap of SemiFL tends to be zero.
For FL, as $N_f + N_c\! \rightarrow \! N$, we have $\frac{\tilde{\rho}_2}{1-\tilde{\rho}_1} \! \rightarrow \! 0$, i.e., the optimality gap of FL tends to be zero, which is consistent with the theorem established in~\cite{Chen2021A}.
For CL, as $N_f\!+\!N_c \! \rightarrow \! N$, we have $\frac{\hat{\rho}_2}{1-\hat{\rho}_1} \! \rightarrow \! 0$, i.e., the optimality gap of CL tends to be zero.

\subsection{Problem Formulation}
\label{problem_formulation}
In the following, we aim to minimize the optimality gap of SemiFL by optimizing the transmit power allocation coefficients $\{p_{t,f,k}\}$ and $\{p_{t,c,k}\}$, as well as the aggregation beamformer $\{{\bf{b}}_t\}$ and decoding beamformers $\{{\bf{f}}_{t,k}\}$.

The transmit power of the $k$-th device is constrained by
\begin{align}
	\label{power_constraint}
	\mathbb{E}[|p_{t,f,k}s_{t,k,q}+p_{t,c,k}d_{t,k,q}|^2]&=\frac{N^2_{f,k}}{N^2_f}\left|p_{t,f,k}\right|^2+\left|p_{t,c,k}\right|^2 \notag \\
	&\le P_{\max},~\forall k \in \mathcal{K},~\forall t,
\end{align}
where $P_{\max}$ is the maximum transmit power at each device.
Suppose that each data sample has $m$ bits.
The $k$-th device should complete the data transmission by the end of each communication round.
Thus, the communication latency of the $k$-th device should not exceed the maximum allowable latency $T_c$, given by
\begin{align}
	\label{data_rate_constraint}
	  \frac{mN_{c,k}}{Wb_1{\log}_2 \left( 1 + \frac{\gamma_{t,k}}{b_2} \right)} \le T_c,~\forall k \in \mathcal{K},~\forall t,
\end{align}
where $0<b_1<1$ and $b_2>1$ are two constants standing for the rate adjustment and the SINR gap~\cite{Lu2021Distortion,Mazzotti2012Multiuser}, respectively, and $W$ is the bandwidth.

To improve the convergence of SemiFL, we minimize the optimality gap $\psi_T^{\rm SemiFL}\left( \{ p_{f,k} \}, {\bf{b}} \right)$ in Theorem~\ref{theorem_1} by jointly optimizing the transmit power allocation and receive beamformers.
Since $\rho_1$ and $\rho_2$ in (\ref{convergence_bound}) are free from the impact of transceiver design, it is equivalent to minimizing $\sum \nolimits_{t=1}^{T} {\rho_1^{T-t} \varphi_t\left( \{ p_{f,k} \}, {\bf{b}} \right)}$.
Therefore, we formulate the optimization problem as
\begin{subequations}
	\label{p_1}
	\begin{eqnarray}
		&\mathop {\min }\limits_{\{p_{t,f,k}\},\{p_{t,c,k}\},\atop{\{{\bf{b}}_t\},\{{\bf{f}}_{t,k}\}}} & \sum\limits_{t=1}^{T} {\rho_1^{T-t} \varphi_t\left( \{ p_{f,k} \}, {\bf{b}} \right)} \\
		\label{MSE_constraint}
		&{\rm s.t.}& {\rm MSE}_t \le \epsilon, \forall t, \\ 
		&{}& \text{(\ref{power_constraint})},~\text{(\ref{data_rate_constraint})}, \notag
	\end{eqnarray}
\end{subequations}
where $\epsilon$ denotes the MSE tolerance. 
Constraint (\ref{power_constraint}) restricts the transmit power of devices.
Constraint (\ref{data_rate_constraint}) specifies the latency requirement of NOMA-based data uploading.
Constraint (\ref{MSE_constraint}) limits the distortion of AirComp-based gradient aggregation.

Although problem (\ref{p_1}) is an optimization problem over $T$ rounds, we observe that both the objective and constraints corresponding to different rounds are independent.
As a result, problem (\ref{p_1}) can be equivalently decomposed into $T$ one-round optimization problems~\cite{Huang2022Wireless}.
We turn to solve the decomposed problems independently for each round.
Note that the problem decomposition empowers SemiFL with the practicability to be implemented over wireless channels varying between rounds.
After removing constant terms in the objective function, the problem in an arbitrary round is rewritten as follows, where the subscript $t$ is omitted.
\begin{subequations}
	\label{p_2}
	\begin{eqnarray}
		\label{original_objective}
		\hspace{-1.1 cm}&\mathop {\min }\limits_{\{p_{c,k}\},\{p_{f,k}\},\atop{{\bf{b}},\{{\bf{f}}_{k}\}}} &\hspace{-0.2 cm} \! \sum\limits_{k=1}^K \! { \frac{4KN_{f,k}^2}{(N_f\!+\!N_c)^2}\! \left| 1\! -\! p_{f,k} {\bf{b}}^{\rm H} {\bf{h}}_{k} \right|^2} \!\!\!+\!\! \frac{N_f^2\sigma^2\|{\bf{b}}\|^2}{(N_f\!+\!N_c)^2} \\
		\label{original_constraint_1}
		\hspace{-1.1 cm}&{\rm s.t.}&\hspace{-0.2 cm} \frac{N^2_{f,k}}{N^2_f}\left|p_{f,k}\right|^2+\left|p_{c,k}\right|^2 \le P_{\max},~\forall k \in \mathcal{K}, \\
		\label{original_constraint_2}
		\hspace{-1.1 cm}&{}&\hspace{-0.2 cm} \gamma_k \ge \gamma_{\min,k},~\forall k \in \mathcal{K}, \\
		\label{original_constraint_3}
		\hspace{-1.1 cm}&{}&\hspace{-0.2 cm} {\rm MSE} \le \epsilon, 
	\end{eqnarray}
\end{subequations}
where $\gamma_{\min,k}=b_2(2^{({mN_{c,k}}/{b_1WT_c})}-1), \forall k \in \mathcal{K}$.
Problem (\ref{p_2}) is non-convex due to the non-convexity of (\ref{original_constraint_2}) and the concave terms in (\ref{original_constraint_3}).
To make problem (\ref{p_2}) tractable, we propose a two-stage algorithm to solve it in the next section.

\section{Transceiver Optimization}
\label{optimization_of_the_transmit_power_control_policies_and_receive_beamformers}
Since the objective function of the formulated problem (\ref{p_2}) is independent of decoding beamformers $\{{\bf{f}}_k\}$, we decompose problem (\ref{p_2}) into two subprblems and propose a two-stage algorithm to solve them for each communication round.
We first jointly optimize the aggregation beamformer and the transmit power allocation coefficients. 
Then, the decoding beamformers are obtained by employing SCA to solve the decomposed $K$ independent problems, where optimal solutions in closed forms are provided by solving KKT conditions.

\subsection{Optimizing Aggregation Beamformer and Transmit Power}
\label{transmit_power_control_policies_and_b}
Given decoding beamformers $\{{\bf{f}}_k\}$, the first subproblem aims to jointly optimize the power allocation coefficients $\{p_{f,k}\}$ and $\{p_{c,k}\}$ as well as the aggregation beamformer $\bf{b}$, given by
\begin{eqnarray}
	\label{p_3}
	\hspace{-1.1 cm}&\mathop {\min }\limits_{\{p_{f,k}\},\atop{\{p_{c,k}\},
			{\bf{b}}}} & \hspace{-0.1 cm} \! \sum\nolimits_{k=1}^K \! { \frac{4KN_{f,k}^2}{(N_f \!+\!\! N_c)^2} \! \left| 1 \!\! - \!\! p_{f,k} {\bf{b}}^{\rm H} {\bf{h}}_{k} \right|^2} \hspace{-0.2 cm} + \! \frac{N_f^2\sigma^2\|{\bf{b}}\|^2}{(N_f \!+\!\! N_c)^2} \\
	\hspace{-1.1 cm}&{\rm s.t.}& \hspace{-0.1 cm} \text{(\ref{original_constraint_1})}-\text{(\ref{original_constraint_3})}. \notag
\end{eqnarray}
Problem (\ref{p_3}) is still non-convex due to the coupling of optimization variables in the objective function and constraints (\ref{original_constraint_2}) and (\ref{original_constraint_3}).
We further decouple problem (\ref{p_3}) into two subproblems regarding the aggregation beamformer and the transmit power allocation coefficients, respectively.

\subsubsection{Subproblem of Aggregation Beamformer Design}
Given decoding beamformers $\{{\bf{f}}_k\}$ and the transmit power allocation coefficients $\{p_{f,k}\}$ and $\{p_{c,k}\}$, the subproblem w.r.t. the aggregation beamformer $\bf{b}$ is given by
\begin{subequations}
	\label{p_5}
	\begin{eqnarray}
		\hspace{-1.1 cm}&\mathop {\min }\limits_{{\bf{b}}} &\hspace{-0.1 cm} {\bf{b}}^{\rm H} {\bf{A}}_0 {\bf{b}} \!-\! 2{\mathop{\rm Re}\nolimits} \left\{ {\bf{b}}^{\rm H} \sum\limits_{k=1}^K {\frac{4KN_{f,k}^2p_{f,k}}{(N_f+N_c)^2}{\bf{h}}_k} \right\} \\
		\label{beamformer_b_constraint_1}
		\hspace{-1.1 cm}&{\rm s.t.}&\hspace{-0.1 cm} {\bf{b}}^{\rm H} {\bf{A}}_{1} {\bf{b}} \!-\! 2{\mathop{\rm Re}\nolimits} \left\{ {\bf{b}}^{\rm H} \sum\limits_{k=1}^K {\frac{N_{f,k}^2p_{f,k}}{N_f^2}{\bf{h}}_k} \right\} \!+\! \iota \!-\! \epsilon \le 0,
	\end{eqnarray}
\end{subequations}
where $\iota=\sum\nolimits_{k=1}^K {\frac{N_{f,k}^2}{N_f^2}}$, and ${\bf{A}}_0$ and ${\bf{A}}_1$ are given by
\begin{align}
	\label{auxiliary_matrice_A0}
	\hspace{-0.2 cm}{\bf{A}}_0=& \sum\limits_{k=1}^K { \frac{4KN_{f,k}^2|p_{f,k}|^2}{(N_f+N_c)^2} {\bf{h}}_k {\bf{h}}_k^{\rm H} + \frac{N_f^2\sigma^2}{(N_f+N_c)^2}{\bf{I}}_{N_r}},\\
	\label{auxiliary_matrice_A1}
	\hspace{-0.2 cm}{\bf{A}}_1=&\sum\limits_{k=1}^K {\frac{N_{f,k}^2|p_{f,k}|^2}{N_f^2}{\bf{h}}_k{\bf{h}}_k^{\rm H}} + \sigma^2 {\bf{I}}_{N_r}.
\end{align}
Problem (\ref{p_5}) is convex w.r.t. ${\bf{b}}$ because of the positive semidefinite matrices ${\bf{A}}_0$ and ${\bf{A}}_1$.
Therefore, the problem can be numerically solved using standard optimization toolboxes, such as CVX~\cite{Grant2014CVX}.
In a special case where the BS has a single antenna, we obtain the optimal aggregation beamformer in the following Lemma by solving KKT conditions.
\begin{lemma}
	\label{lemma_3}
	When the BS is single-antenna, the optimal aggregation beamformer is given by
	\begin{align}
		\label{b_solution_single_antenna}
		b^*=\frac{\hat{h}_1}{\omega_1}+{\sqrt{|\frac{\hat{h}_1}{\omega_1}|^2-\frac{\iota-\epsilon}{\omega_1}}}e^{i\angle{(\omega_1\hat{h}_0-\omega_0\hat{h}_1)}},
	\end{align}
	where $\omega_0=\sum\nolimits^K_{k=1}\!{\frac{4KN_{f,k}^2|p_{f,k}|^2|{h}_k|^2}{(N_f+\!N_c)^2}  + \frac{N_f^2\sigma^2}{(N_f+\!N_c)^2}}$, $\omega_1=\sum\nolimits_{k=1}^K{\frac{N_{f,k}^2|p_{f,k}|^2|h_k|^2}{N_f^2}}+\sigma^2$, $\hat{h}_0=\sum\nolimits_{k=1}^K{\frac{4KN_{f,k}^2|p_{f,k}|^2{h}_k}{(N_f+N_c)^2}}$, $\hat{h}_1=\sum\nolimits_{k=1}^K{\frac{N_{f,k}^2p_{f,k}h_k}{N_f^2}}$, and $i=\sqrt{-1}$.
\end{lemma}
\begin{IEEEproof}
	Please refer to Appendix~\ref{proof_of_lemma_3}.
\end{IEEEproof}

\subsubsection{Subproblem of Transmit Power Allocation}
\label{subproblem_transmit_power_control_policies}
Given the aggregation and decoding beamformers $\bf{b}$ and $\{{\bf{f}}_k\}$, the subproblem of transmit power allocation coefficients reduces to
\begin{eqnarray}
	\label{p_6}
	\hspace{-0.5 cm}&\mathop {\min }\limits_{\{p_{f,k}\},\{p_{c,k}\}} & \sum\limits_{k=1}^K { \frac{4KN_{f,k}^2}{(N_f+N_c)^2} \left| 1 - p_{f,k} {\bf{b}}^{\rm H} {\bf{h}}_{k} \right|^2} \\
	\hspace{-0.5 cm}&{\rm s.t.}& \text{(\ref{original_constraint_1})}-\text{(\ref{original_constraint_3})}, \notag
\end{eqnarray}
which is also non-convex due to the indefinite Hessian matrices of (\ref{original_constraint_2}) and (\ref{original_constraint_3}).

With reference to~\cite{Qin2021Over}, we have ${\mathop{\rm Re}\nolimits} \{p_{f,k} {\bf{b}}^{\rm H} {\bf{h}}_{k}\} \le |p_{f,k}||{\bf{b}}^{\rm H} {\bf{h}}_{k}|$.
As a result, it is obtained that $| 1 - p_{f,k} {\bf{b}}^{\rm H} {\bf{h}}_{k} |^2 = 1 + |p_{f,k}|^2|{\bf{b}}^{\rm H} {\bf{h}}_{k}|^2 - 2{\mathop{\rm Re}\nolimits} \{p_{f,k} {\bf{b}}^{\rm H} {\bf{h}}_{k}\} \ge (1-|p_{f,k}||{\bf{b}}^{\rm H} {\bf{h}}_{k}|)^2$, where the equality holds if $\angle{p_{f,k}}+\angle({\bf{b}}^{\rm H} {\bf{h}}_{k})=0$.
Therefore, we determine the angles of $\{p_{f,k}\}$ as
\begin{align}
	\label{angle_of_pk}
	\angle{p_{f,k}}=-\angle({\bf{b}}^{\rm H} {\bf{h}}_{k}),~\forall k \in \mathcal{K}.
\end{align}
Consider that the angles of $\{p_{c,k}\}$ are independent of problem (\ref{p_6}).
For simplicity, we determine the angles of $\{p_{c,k}\}$ by
\begin{align}
	\label{angle_of_p_ck}
	\angle p_{c,k}=0,~\forall k \in \mathcal{K}.
\end{align}

Based on the obtained angles, we perform variable substitutions by letting $\alpha_{k}=|p_{f,k}|,~\forall k \in \mathcal{K}$ and $\beta_{k}=|p_{c,k}|^2,~\forall k \in \mathcal{K}$.
Consequently, problem (\ref{p_6}) is rewritten as
\begin{subequations}
	\label{p_7}
	\begin{eqnarray}
		\hspace{-1 cm}&\mathop {\min }\limits_{\{\alpha_{k}\},\atop{\{\beta_{k}\}}} \hspace{-0.3 cm}& \sum\nolimits_{k=1}^K { \frac{4KN_{f,k}^2}{(N_f+N_c)^2} \left( 1 - \alpha_{k} |{\bf{b}}^{\rm H} {\bf{h}}_{k}| \right)^2} \\
		\label{power_constraint_1}
		\hspace{-1 cm}&{\rm s.t.}\hspace{-0.2 cm}& \frac{N^2_{f,k}}{N^2_f}\alpha_{k}^2+\beta_{k} - P_{\max} \le 0, \forall k \in \mathcal{K}, \\
		\label{power_constraint_2}
		\hspace{-1 cm}&{}\hspace{-0.2 cm}& -\beta_{k}\left|{\bf{f}}_{k}^{\rm H}{\bf{h}}_{k}\right|^2\! +\! \gamma_{\min,k}( \sum \nolimits_{k'=1}^K \! {\frac{N_{f,k'}^2}{N_f^2}\alpha_{k'}^2|{\bf{f}}_{k}^{\rm H}{{\bf{h}}_{k'}}|^2} \notag \\
		\hspace{-1 cm}&{}\hspace{-0.2 cm}&+\sum\nolimits_{{k'}=1,{k' \neq k}}^K {\beta_{k'}|{\bf{f}}_{k}^{\rm H}{\bf{h}}_{k'}|^2} \!\!+\! \sigma^2\!\|{\bf{f}}_k\|^2) \! \le \! 0, \forall k \in \mathcal{K}, \\
		\label{power_constraint_3}
		\hspace{-1 cm}&{}\hspace{-0.2 cm}& \sum\nolimits_{k=1}^K {\frac{N_{f,k}^2}{N_f^2}(1\!-\!\alpha_{k}|{\bf{b}}^{\rm H}{\bf{h}}_k|)^2} \!+\! \|{\bf{b}}\|^2\sigma^2 \!-\! \epsilon \! \le \! 0, \\
		\hspace{-1 cm}&{}\hspace{-0.2 cm}& \beta_{k} \ge 0, \alpha_{k} \ge 0, \forall k \in \mathcal{K}.
	\end{eqnarray}
\end{subequations}
Due to positive semidefinite Hessian matrices of the objective and constraints, problem (\ref{p_7}) is jointly convex w.r.t. $\{\alpha_{k}\}$ and $\{\beta_{k}\}$, and thus can be numerically solved.
Finally, the transmit power allocation coefficients are recovered by
\begin{align}
	\label{solution_p_fk}
	&p^*_{f,k}=\alpha_{k}e^{i\angle{p_{f,k}}},~\forall k \in \mathcal{K},\\
	\label{solution_p_ck}
	&p^*_{c,k}=\sqrt{\beta_{k}}e^{i\angle p_{c,k}},~\forall k \in \mathcal{K}.
\end{align}

\subsection{Optimizing Decoding Beamformers}
\label{receive_beamformers_f_k}
Given the power coefficients $\{p_{f,k}\}$ and $\{p_{c,k}\}$, as well as the aggregation beamformer $\bf{b}$, the second subproblem attempts to find feasible decoding beamformers $\{{\bf{f}}_k\}$, which is rewritten as
\begin{subequations}
	\label{p_8}
	\begin{eqnarray}
		&\mathop {\text{find} }\limits_{\{{\bf{f}}_{k}\}} & \{{\bf{f}}_k\} \\
		&{\rm s.t.}& {\bf{f}}_k^{\rm H} {\bf{A}}_{2,k} {\bf{f}}_k \le 0, \forall k \in \mathcal{K},
	\end{eqnarray}
\end{subequations}
where ${\bf{A}}_{2,k}$ is given by
\begin{align}
	\label{auxiliary_matrice_A2}
	&{\bf{A}}_{2,k}=-|p_{c,k}|^2{\bf{h}}_k{\bf{h}}_k^{\rm H} + \gamma_{\min,k}  \left( \sum\nolimits_{k' = 1}^K {\frac{N_{f,k'}^2}{N_f^2}|p_{f,k'}|^2{\bf{h}}_{k'}{\bf{h}}_{k'}^{\rm H}}\right. \notag \\
	&\left.+ \sum\nolimits_{k'= 1, {k' \neq k}}^K  {|p_{c,k'}|^2{\bf{h}}_{k'}{\bf{h}}_{k'}^{\rm H}}  + \sigma^2 {\bf{I}}_{N_r} \right), \forall k \in \mathcal{K}.
\end{align}

Considering the independence of constraints among devices, we decompose (\ref{p_8}) into $K$ independent problems w.r.t. each device.
With the aim of increasing the individual data rate
\footnote{All devices ought to complete their data sample transmissions within the same time duration when NOMA is employed~\cite{Fang2020Optimal}.
However, increasing the individual data rate might lead to misaligned transmission latency among devices.
To address this issue, the number of bits used to represent a data sample at each device, i.e., $m$ bits, should be carefully adjusted in terms of the obtained individual data rate to align the data sample transmission latency of different devices.}
, we introduce an auxiliary variable $\nu_k \le 0$ to transform the problem of the $k$-th device as follows~\cite{Liu2021Intelligent}:
\begin{subequations}
	\label{p_9}
	\begin{eqnarray}
		&\mathop {\min }\limits_{{\bf{f}}_{k},\nu_k\le0} & \nu_k \\
		\label{f_constraint_1}
		&{\rm s.t.}& {\bf{f}}_k^{\rm H} {\bf{A}}_{2,k} {\bf{f}}_k - \nu_k \le 0,
	\end{eqnarray}
\end{subequations}
which is a non-convex problem due to the indefinite matrix ${\bf{A}}_{2,k}$ in constraint (\ref{f_constraint_1}).

We employ the SCA method to solve problem (\ref{p_9}).
The surrogate function of ${\bf{f}}_k^{\rm H} {\bf{A}}_{2,k} {\bf{f}}_k$, i.e., $g({{\bf{f}}}_k|{{\bf{f}}}^{(n)}_k)$, is created as follows~\cite{Sun2017Majorization}:
\begin{align}
	\label{surrogate_A_2}
	g({{\bf{f}}}_k|{{\bf{f}}}^{(n)}_k) = &{\bf{f}}^{\rm H}_k{\bf{M}}_k{\bf{f}}_k + 2 {\mathop{\rm Re}\nolimits} \{ {\bf{f}}^{\rm H}_k({\bf{A}}_{2,k}-{\bf{M}}_k){\bf{f}}^{(n)}_k \} \notag \\
	&+ ({\bf{f}}^{(n)}_k)^{\rm H}({\bf{M}}_k-{\bf{A}}_{2,k}){\bf{f}}^{(n)}_k,~\forall k \in \mathcal{K},
\end{align}
where the matrix ${\bf{M}}_k$ satisfies ${\bf{M}}_k \succeq {\bf{A}}_{2,k},\forall k \in \mathcal{K}$, and ${{\bf{f}}}^{(n)}_k$ is the result obtained at the $n$-th iteration of SCA.
By substituting (\ref{surrogate_A_2}) into (\ref{f_constraint_1}), we convexify problem (\ref{p_9}) as
\begin{subequations}
	\label{p_10}
	\begin{eqnarray}
		&\mathop {\min }\limits_{{\bf{f}}_{k},\nu_k\le0} & \nu_k \\
		&{\rm s.t.}& g({{\bf{f}}}_k|{{\bf{f}}}^{(n)}_k) - \nu_k \le 0,
	\end{eqnarray}
\end{subequations}
By solving KKT conditions, the close-form optimal solutions to problem (\ref{p_10}) is provided  in the following lemma.
\begin{lemma}
	\label{lemma_4}
	The optimal solutions to problem (\ref{p_10}) are given by
	\begin{align}
		\label{f_k_solution}
		&{\bf{f}}^*_k={\bf{M}}^{-1}_k({\bf{M}}_k-{\bf{A}}_{2,k}){\bf{f}}^{(n)}_k, \forall k \in \mathcal{K} \\
		\label{nu_k_solution}
		&\nu^*_k=({\bf{f}}^{(n)}_k)^{\rm H}{\bf{A}}_{2,k}{\bf{M}}^{-1}_k({\bf{M}}_k-{\bf{A}}_{2,k}){\bf{f}}^{(n)}_k, \forall k \in \mathcal{K}.
	\end{align}
\end{lemma}
\begin{IEEEproof}
	Please refer to Appendix~\ref{proof_of_lemma_4}.
\end{IEEEproof}

\begin{algorithm}[t]
	\caption{A Two-Stage Algorithm for Solving (\ref{p_2})}
	\label{algorithm_1}
	\begin{algorithmic}[1]
		\STATE \textbf{Input: }Feasible solutions $(\{p_{f,k}^{(0)}\}, \{p_{c,k}^{(0)}\},{\bf{b}}^{(0)}\!, \{{\bf{f}}_{k}^{(0)}\})$, the maximum number of iterations $N$, the convergence accuracy $\varepsilon$, $n=0$, and $n'=0$.
		\REPEAT
		\STATE Update $n \leftarrow n+1$.
		\STATE Given $\{p_{f,k}^{(n-1)}\}$,$\{p_{c,k}^{(n-1)}\}$ and $\{{\bf{f}}^{(0)}_k \}$, obtain ${\bf{b}}^{(n)}$ by solving problem (\ref{p_5}).
		\STATE Given ${\bf{b}}^{(n)}$ and $\{{\bf{f}}_{k}^{(0)} \}$, calculate $\angle p_{f,k}^{(n)}, \forall k \in \mathcal{K}$ by (\ref{angle_of_pk}) and $\angle p_{c,k}^{(n)}, \forall k \in \mathcal{K}$ by (\ref{angle_of_p_ck}).
		\STATE Given ${\bf{b}}^{(n)}$, $\{{\bf{f}}_{k}^{(0)}\}$, obtain $\{\alpha_{k}^{(n)}\}$ and $\{\beta_{k}^{(n)}\}$ by solving problem (\ref{p_7}).
		\STATE Recover $p_{f,k}^{(n)}, \forall k \! \in \! \mathcal{K}$ by (\ref{solution_p_fk}) and $p_{c,k}^{(n)}, \forall k \! \in \mathcal{K}$ by (\ref{solution_p_ck}).
		\UNTIL $n \ge {N}$ or $ \frac{|{U}^{(n)}-{U}^{(n-1)}|}{|{U}^{(n)}|} \le \varepsilon$
		\REPEAT
		\STATE Update $n' \leftarrow n'+1$.
		\STATE Given $\{p_{f,k}^{(n)}\}$, $\{p_{c,k}^{(n)}\}$, ${\bf{b}}^{(n)}$ and $\{{\bf{f}}^{(n'-1)}_k \}$, obtain ${\bf{f}}_k^{(n')},\forall k \in \mathcal{K}$ by (\ref{f_k_solution}) and ${\nu}^{(n')}_k, \forall k \in \mathcal{K}$ by (\ref{nu_k_solution}).
		\UNTIL $n' \ge {N}$ or $ \frac{|{\nu_k}^{(n')}-{\nu_k}^{(n'-1)}|}{|{\nu_k}^{(n')}|} \le \varepsilon$
		\STATE \textbf{Output: }The solution $(\{p_{f,k}^{(n)}\}$, $\{p_{c,k}^{(n)}\}$, ${\bf{b}}^{(n)}$, $\{{\bf{f}}_{k}^{(n')}\})$.
	\end{algorithmic}
\end{algorithm}

\subsection{Algorithm, Convergence and Complexity}
The proposed two-stage algorithm for solving problem (\ref{p_2}) is summarized in Algorithm~\ref{algorithm_1}, where the superscript $n$ denotes the $n$-th iteration and $U$ denotes the value of (\ref{original_objective}).
In light of the non-increase and non-negativity of objective (\ref{original_objective}) over iterations, the convergence of Algorithm~\ref{algorithm_1} can be conformed based on the Monotone Bounded Theorem~\cite{Boyd2004Convex}.
By adopting the standard interior-point (SIP) method when invoking CVX, the worst-case complexity of Algorithm~\ref{algorithm_1} is given by $\mathcal{O} (NN_1N_r^3+8NN_2K^3+KN)$, where $N_1$ and $N_2$ are the permitted maximum iterations of SIP for problems (\ref{p_5}) and (\ref{p_7}), respectively.
Specifically, $\mathcal{O} (N_1 N_r^3)$ and $\mathcal{O} (8 N_2 K^3)$ present the complexities of solving problems (\ref{p_5}) and (\ref{p_7}), respectively.
In terms of the closed-form solution in (\ref{f_k_solution}), the complexity for solving ${\bf{f}}_{k}$ is $\mathcal{O}(1)$.

\section{Simulation Results}
\label{simulation_results}

\subsection{Simulation Setup}
\label{simulation_setup}
We consider a SemiFL system with a radius of $100$ m, wherein $K=10$ devices are randomly located.
The BS equipped with $N_r=16$ antennas is located at the coordinate $(0,0,10)$ m. 
The large-scale fading coincides with that in~\cite{Ni2021Resource}, and consider Rician factor $\kappa=2$ for the small-scale fading.
The transmission bandwidth is $W=5$ MHz. 
The noise power is $\sigma^2=-80$ dBm, and the maximum transmit power is $P_{\max}=30$ dBm.
The rate adjustment and the SINR gap are set as $b_1=0.905$ and $b_2=1.34$~\cite{Lu2021Distortion}, respectively.
Other parameters are set as $\epsilon=0.5$, $N=200$, and $\varepsilon=0.01$.

We verify the performance of SemiFL by conducting classification experiments on the MNIST and CIFAR-10 datasets~\cite{Cui2022AFast}:
\begin{enumerate}
	\item For the MNIST dataset, each data sample comprises a $28 \times 28$ gray-scale image and a $10$-dimensional label.
	Representing each entry by $16$ bits, we set $m\!=\!(28\!\times\!28+10)\times16\!=\!12,704$ bits.
	The global model is a fully-connected multilayer perceptron (MLP) with a $50$-neuron hidden layer, which has $Q=39,760$ parameters in total.
	When training the MLP using SemiFL, the MSE loss function is adopted and the learning rate is $\eta=0.01$.
	\item For the CIFAR-10 dataset, each data sample consists of a $32 \times 32 \times 3$ color image and a $10$-dimensional label, which contains $m=(32\times32\times3+10)\times16=49,312$ bits.
	We utilize a $9$-layer convolutional neural network (CNN) with $Q=116,906$ parameters as the global model for classification.
	There are three convolutional layers with ReLU activation, three max pooling layers of size $3\times3$, two fully-connected layers with ReLU activation, and one softmax output layer in the CNN.
	The three convoltion layers contains $32$, $32$, and $64$ kernels of size $5\! \times\! 5$, respectively.
	There are $64$ and $10$ neurons in the two fully connected layers, respectively.
	We adopt the cross-entropy loss function and set the learning rate as $\eta\!=\!0.1$ to train the CNN using SemiFL.
\end{enumerate}
For the above two classification experiments, we consider $T=1,000$ training rounds with the maximum communication allowable latency $T_c=500$ ms. Each device independently and randomly draws $N_{f,k}\!+\!N_{c,k}\!=\!24$ data samples from the corresponding training set in each round. 
Then, $N_{f,k}\!=\!16$ samples are retained locally for FL, and $N_{c,k}\!=\!8$ samples are uploaded to the BS for CL.
The classification accuracy is evaluated on the entire test set.

\subsection{Evaluation of Communication Metrics}
\label{wireless-related_performance}

\begin{figure*}
	\centering
	\begin{minipage}[t]{0.49 \textwidth}
		\centering
		\includegraphics[width=3.45 in]{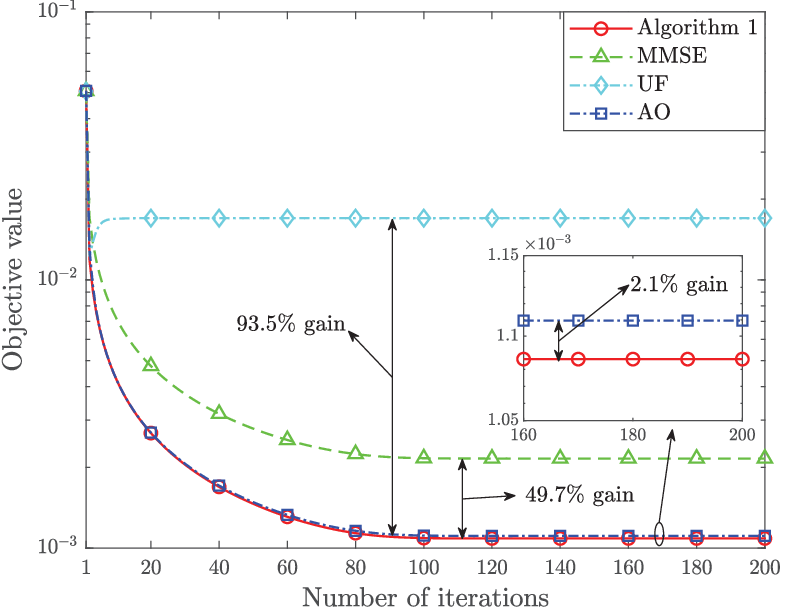}
		\caption{Comparison of convergence behaviors.}
		\label{simulation_convergence}
	\end{minipage}
	\begin{minipage}[t]{0.49 \textwidth}
		\centering
		\includegraphics[width=3.35 in]{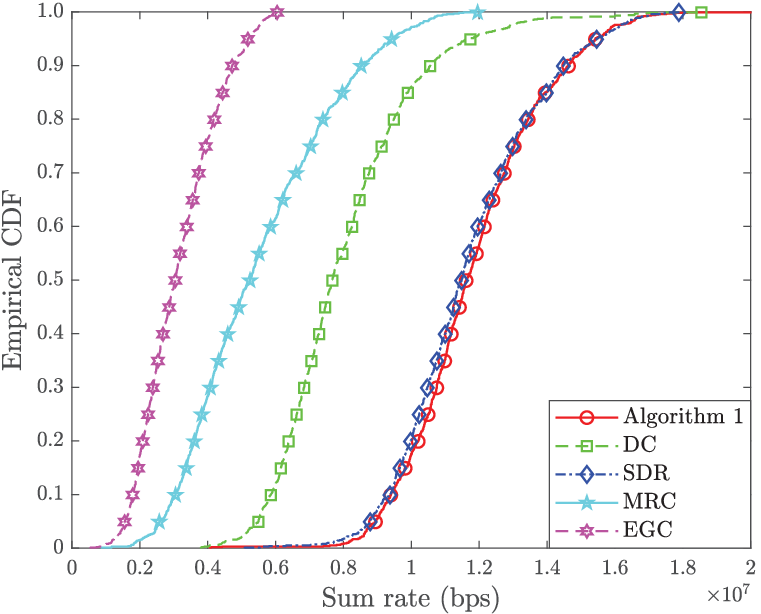}
		\caption{Empirical CDF of the sum rate.}
		\label{simulation_EDF}
	\end{minipage}\\
	\vspace{+0.2 cm}
	\begin{minipage}[t]{0.49 \textwidth}
		\centering
		\includegraphics[width=3.45 in]{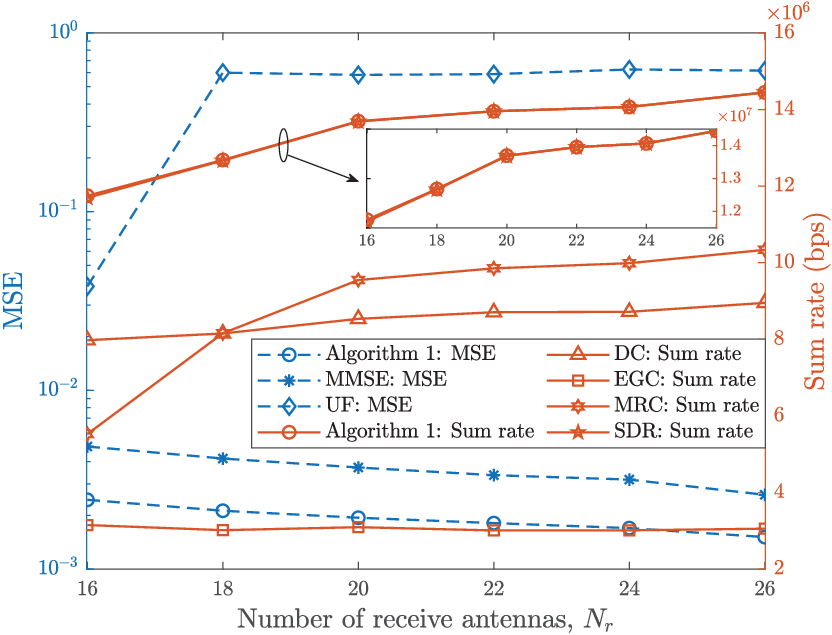}
		\caption{MSE and sum rate versus the number of receive antennas under different schemes.}
		\label{simulation_MSE_sum_rate}
	\end{minipage}
	\begin{minipage}[t]{0.49 \textwidth}
		\centering
		\includegraphics[width=3.45 in]{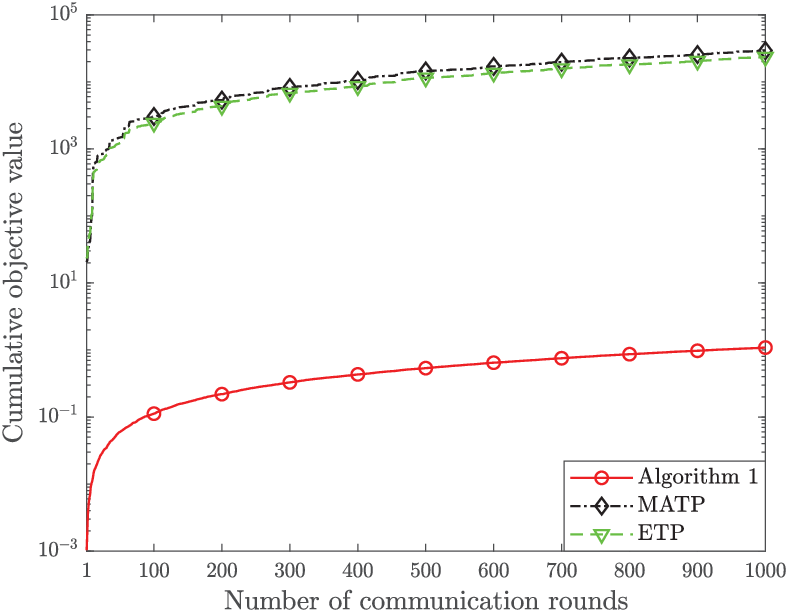}
		\caption{Cumulative objective value versus the number of communication rounds under different schemes.}
		\label{simulation_cumulative_obj_vs_rounds}
	\end{minipage}
\end{figure*}

In Fig.~\ref{simulation_convergence}, we plot the convergence behavior of Algorithm~\ref{algorithm_1} in comparison with three benchmarks, including:
i) the BS configures the aggregation beamformer as a minimum MSE (MMSE) receiver~\cite{Mansour2021Low};
ii) devices employ uniform-forcing (UF) transmitters~\cite{Chen2018Uniform};
iii) the aggregation beamformer and the transmission power allocation coefficients are solved using alternating optimization (AO)~\cite{Ni2021Federated}.
It is seen that the objective value of (\ref{original_objective}) monotonously decreases with iterations and finally reaches the stationary point.
In particular, Algorithm~\ref{algorithm_1} effectively reduces the optimality gap and outperforms the benchmarks by converging to the lowest objective value.

In Fig.~\ref{simulation_EDF}, we plot the empirical cumulative distribution function (CDF) of the sum rate for $1,000$ trials.
We employ the sum rate as another metric to evaluate the performance of data transmission, which is defined by the sum rates of each device, i.e., $\sum_{k=1}^{K} Wb_1 \log_{2}\left(1+\frac{\gamma_k}{b_2}\right) $ (bps).
We consider four beamforming schemes as benchmarks:
i) the difference-of-convex-functions (DC)~\cite{Yang2020Federated} method, where matrix lifting is employed to solve ${\bf{f}}_k$ and the rank constraint is approximated by its linearization;
ii) semidefinite relaxation (SDR)~\cite{Luo2010Semidefinite}, where the rank constraint is simply dropped;
iii) maximum ratio combining (MRC)~\cite{Wei2018Joint}, where ${\bf{f}}_k$ is configured as ${\bf{h}}^{\rm H}_{t,k}$;
iv) equal-gain combining (EGC), where ${\bf{f}}_k$ is set as ${\bf{1}}\in\mathbb{C}^{N_r}$.
It is noticed that Algorithm~\ref{algorithm_1} is the right-most among all curves and achieves the highest sum rate. 
This is because SCA circumvents the performance loss due to matrix approximation and decomposition.
Additionally, EGC even dissatisfies the rate request because of the static configuration property.

In Fig.~\ref{simulation_MSE_sum_rate}, we show the impacts of the number of receive antennas on the sum rate and MSE, where the blue dashed curves refer to the MSE performance and the brown solid curves represent the sum rate performance.
It is seen that equipping the BS with more receive antennas results in a lower MSE and a higher sum rate.
Meanwhile, Algorithm~\ref{algorithm_1} outperforms benchmarks in MSE and attains the highest sum rate.
Despite the comparable sum rate of SDR to Algorithm~\ref{algorithm_1}, SDR confronts a quartic complexity regarding $N_r$~\cite{Luo2010Semidefinite}, which is much more time-consuming than the closed-form solution to ${\bf{f}}_k$.
This confirms the practicability of Algorithm~\ref{algorithm_1} in terms of the performance and cost.

In Fig~\ref{simulation_cumulative_obj_vs_rounds}, we plot the cumulative objective value of (\ref{original_objective}) attained by Algorithm~\ref{algorithm_1} and benchmarks, wherein a lower cumulative objective value indicates a better convergence behavior.
To showcase the advantage of the proposed power allocation method, two schemes are considered as benchmarks:
i) maximum available transmit power (MATP), where $p_{f,k}=\sqrt{(N_f^2/N_{f,k}^2)(P_{\max}-|p_{c,k}|^2)}e^{-i\angle{{\bf{b}}^{\rm H}{\bf{h}}_{k}}},\forall k \in \mathcal{K}$;
ii) equal transmit power (ETP), where $p_{f,k}=(N_f/N_{f,k})\sqrt{P_{\max}/2}, \forall k \in \mathcal{K}$, $p_{c,k}=\sqrt{P_{\max}/2}, \forall k \in \mathcal{K}$, ${\bf{b}}$ and $\{{\bf{f}}_k\}$ are obtained by solving problems (\ref{p_5}) and (\ref{p_10}), respectively.
It can be observed that Algorithm~\ref{algorithm_1} significantly outperforms MATP and ETP by achieving a lower cumulative objective value, thereby implying a smaller convergence optimality gap.
This is attributed to the effectiveness of Algorithm~\ref{algorithm_1} in adapting the transmit power of devices to varying channel conditions synthetically, which achieves better aggregation of local gradients for a reduced optimality gap.

\begin{figure*}
	\begin{minipage}[t]{1 \textwidth}
		\centering
		\subfigure[Classification accuracy of training an MLP on the MNIST dataset.]{
			\label{simulation_accuracy_iid}
			\includegraphics[width=0.49 \textwidth]{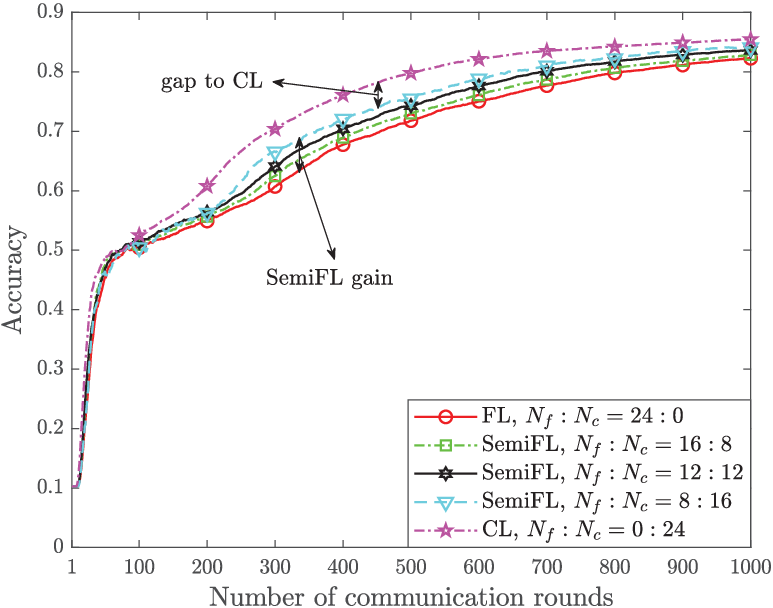}}
		\subfigure[Training loss of training an MLP on the MNIST dataset.]{
			\label{simulation_loss_iid}
			\includegraphics[width=0.49 \textwidth]{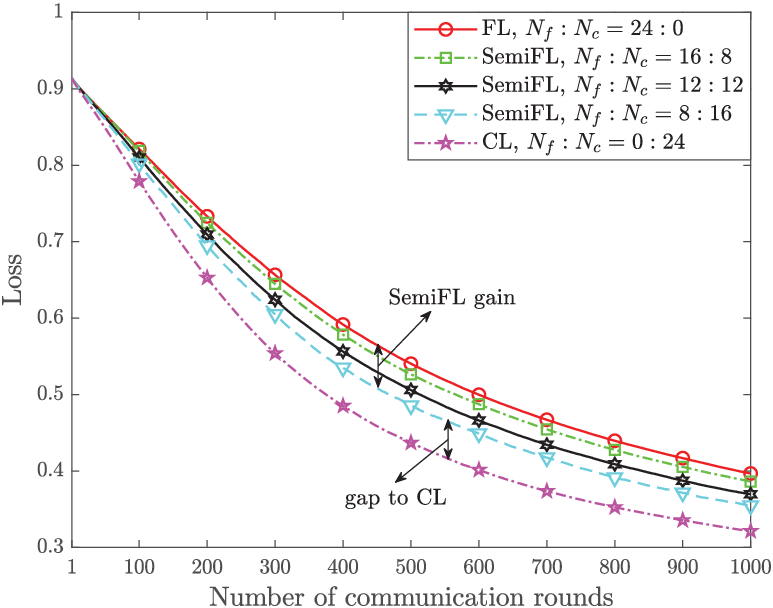}} \\
		\subfigure[Classification accuracy of training a CNN on the CIFAR-10 dataset.]{
			\label{accuracy_CNN_CIFAR-10}
			\includegraphics[width=0.49 \textwidth]{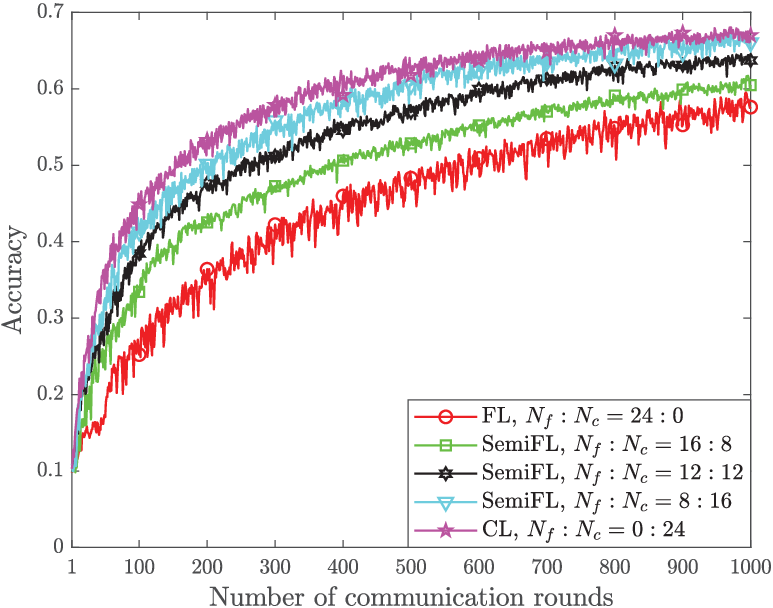}}
		\subfigure[Training loss of training a CNN on the CIFAR-10 dataset.]{
			\label{loss_CNN_CIFAR-10}
			\includegraphics[width=0.49 \textwidth]{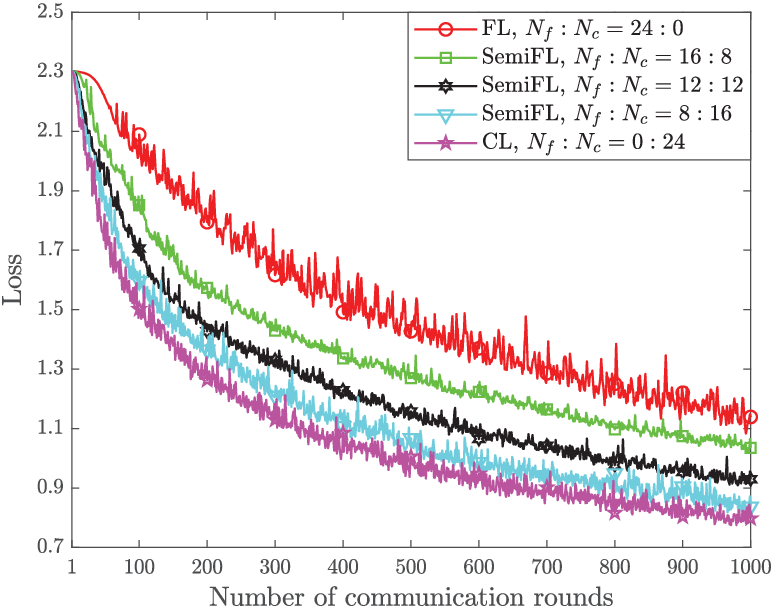}}
		\caption{Learning performance of SemiFL and benchmarks on the MNIST and CIFAR-10 datasets.}
		\label{simulation_iid}
	\end{minipage}
\end{figure*}

\subsection{Classification Experiments on Real-World Datasets}
\label{experiments_on_handwritten_digits_classification}
In this subsection, we examine the effectiveness of SemiFL by conducting classification experiments on the MNIST and CIFAR-10 datasets.
We consider the following five learning benchmarks for comparison:
\begin{enumerate}
	\item FL: The devices merely transmit local gradients to the BS over the same time-frequency resources in each round, i.e., $N_{f,k}=24$ and $N_{c,k}=0,\forall k \in \mathcal{K}$.
	The BS aggregates local gradients over the air and updates the global model with the aggregated gradient.
	\item CL: The devices only transmit all collected data samples in each round to the BS for CL, i.e., $N_{f,k}=0$ and $N_{c,k}=24, \forall k \in \mathcal{K}$.
	The BS calculates the gradient using a batch of its accumulated data samples and updates the global model accordingly.
	\item Hybrid federated and centralized learning (HFCL)~\cite{Elbir2021Hybrid}: The devices are divided into active and passive devices, where the former uploads local models to the BS for FL while the latter sends the entire dataset for CL.
	The learning process does not begin until all passive devices finish uploading datasets to the BS.
	We set half of the devices as active devices and the other half as passive devices.
	\item HFCL with increased computation-per-client (HFCL-ICpC)~\cite{Elbir2021Hybrid}: The only difference between this scheme and HFCL is that, during the data transmission of passive devices, active devices train their own models locally for multiple epochs.
	\item HFCL with sequential data transmission (HFCL-SDT)~\cite{Elbir2021Hybrid}: The only difference between this scheme and HFCL is that, passive devices simultaneously transmit a small part of their data sets to the BS for CL while active devices uploading local models in each round.
	The transmission of passive devices ceases once their entire datasets have been transmitted.
\end{enumerate}
To guarantee fairness, the transmit power coefficients of the active devices and the aggregation beamformer are similarly configured as the first half of the SemiFL devices, and the uploaded data samples from the passive devices are also perfectly decoded like SemiFL.

\begin{figure*}
	\begin{minipage}[t]{0.49 \textwidth}
		\centering
		\includegraphics[width=3.45 in]{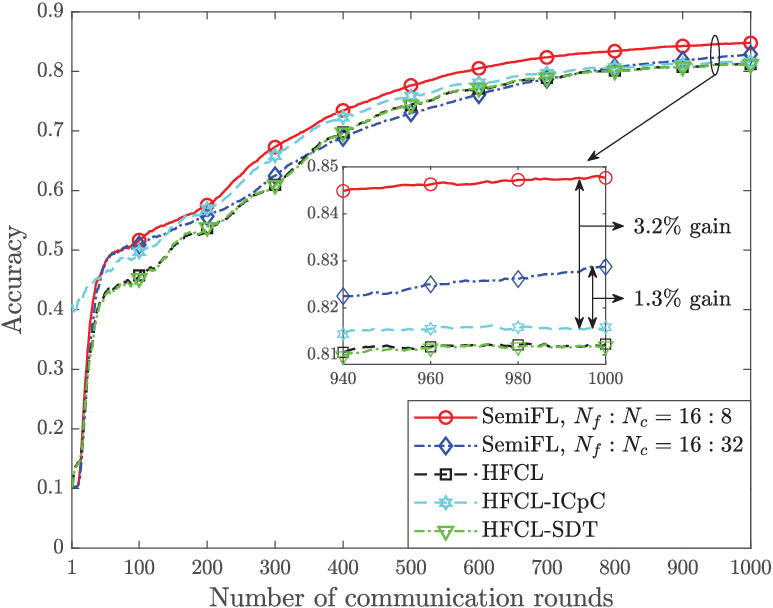}
		\caption{Training process comparison of SemiFL with state-of-the-art learning frameworks on the MNIST dataset.}
		\label{simulation_HFCL_series}
	\end{minipage}
	\begin{minipage}[t]{0.49 \textwidth}
		\centering
		\includegraphics[width=3.45 in]{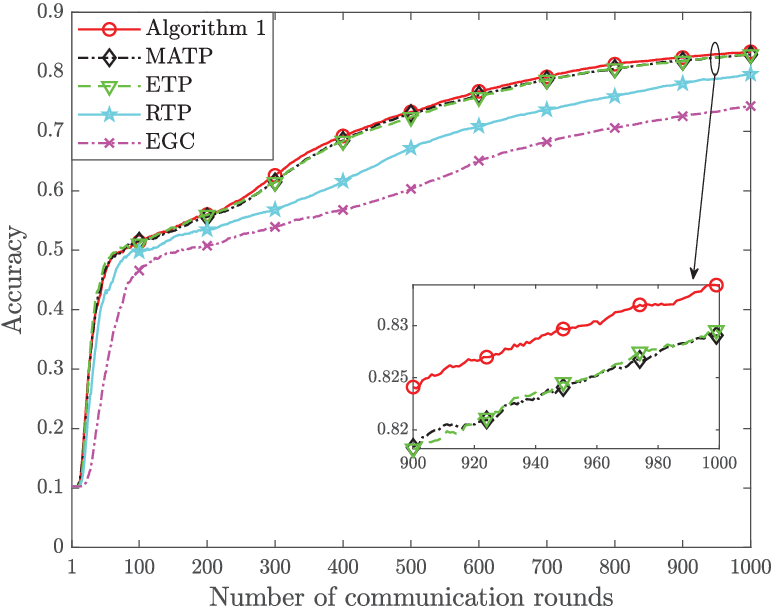}
		\caption{Training process comparison between Algorithm~\ref{algorithm_1} and resource allocation benchmarks on the MINIST dataset.}
		\label{simulation_wireless_benchmarks}
	\end{minipage}
	\vspace{-0.4 cm}
\end{figure*}

In Fig.~\ref{simulation_iid}, we plot the learning performance of SemiFL and benchmarks when training an MLP on the MNIST dataset and a CNN on the CIFAR-10 dataset.
It is worth mentioning that all schemes use the same number of data samples in each round to guarantee the fairness. 
It is observed that SemiFL achieves moderate classification accuracy and training loss between FL and CL in both training settings.
This validates that the proposed SemiFL is a more general learning framework than FL and CL, as demonstrated in Theorem~\ref{theorem_2}.
Moreover, it can be seen that higher accuracy and lower loss can be achieved if more data samples are uploaded to the BS for calculating the CL gradient.
The reason is that the detrimental impact of the wireless channel in aggregation is compensated by the increasing data samples for CL.

In Fig.~\ref{simulation_HFCL_series}, we compare the training process of SemiFL on the MNIST dataset with three state-of-the-art hybrid learning frameworks. 
For comparison fairness, active devices utilize the same batch size as SemiFL devices to train their local models, and the BS employs the gradient descent base on all samples from passive devices.
Despite the higher initial accuracy of HFCL-ICpC due to the local updates in advance, SemiFL eventually outperforms the benchmarks.
It is seen that SemiFL attains $1.3\%$ accuracy gain regarding HFCL-ICpC, which can be enlarged to $3.2\%$ if more samples are dedicated for CL.
This verifies the learning superiority of the proposed SemiFL in terms of the classification accuracy.

In Fig.~\ref{simulation_wireless_benchmarks}, we plot the classification accuracy achieved by Algorithm~\ref{algorithm_1} and resource allocation benchmarks when training an MLP on the MNIST dataset using SemiFL.
Apart from the aforementioned MATP and ETP schemes, the following two resource allocation schemes are employed as benchmarks:
i) random transmit power (RTP), where $p_{f,k}$ is randomly drawn from the available range $(0,\sqrt{(N_f^2/N_{f,k}^2)(P_{\max}-|p_{c,k}|^2)}],\forall k \in \mathcal{K}$, and ${\bf{b}}$ is obtained by solving problem (\ref{p_5});
ii) equal gain combination (EGC), where ${\bf{b}}={\bf{1}} \in \mathbb{C}^{N_r}$.
For comparison fairness, the allocation of other resources except those specified above is the same as Algorithm~\ref{algorithm_1}, and all schemes use the same amount of data samples in each round.
In Fig.~\ref{simulation_wireless_benchmarks}, it is seen that Algorithm~\ref{algorithm_1} outperforms other benchmarks by obtaining higher accuracy.
The result verifies the advantage of Algorithm~\ref{algorithm_1} in terms of classification accuracy, which is credited to the joint optimization of the transceivers.
It is noteworthy that though Algorithm~\ref{algorithm_1} significantly outperforms MATP and ETP in Fig.~\ref{simulation_cumulative_obj_vs_rounds}, the superiority in terms of classification accuracy is less pronounced.
This is because the decrease in convergence optimality gap does not correspond strictly to the same degree of increase in accuracy.
Moreover, one can observe from MATP and ETP schemes that simply exhausting all transmit power for transmitting local gradients can result in reduced accuracy.
This is because poor power control aggravates the distortion of the aggregated signal, which reflects the effectiveness of Algorithm~\ref{algorithm_1} in allocating transmit power again.

\section{Conclusion}
\label{conclusion}
In this paper, we proposed SemiFL, a hybrid learning paradigm in a two-tier framework.
The conventional FL and CL were integrated into a harmonized architecture for improving the learning performance.
To satisfy the distinct transmission requirement of SemiFL, we designed a novel transceiver structure that incorporated NOMA and AirComp to support the JCC principle, which enabled the collaborative uploading of local gradients and data samples.
Then, our theoretical analysis revealed the detrimental effect of poorly configured wireless factors on the convergence of SemiFL, and proved that SemiFL is more general than FL and CL.
In particular, we further extended the convergence analysis to two special cases, and demonstrated that SemiFL over error-free channels could converge to the optimum without any gap once the amount of data samples satisfied specific conditions.
Next, we formulated a non-convex problem to minimize the optimality gap by jointly optimizing the transmitters and receivers.
To solve the problem, we proposed a two-stage algorithm, where closed-form optimal beamformers were provided.
Experiment results on real-world datasets confirmed the theoretical analysis, and illustrated that SemiFL outperformed FL and state-of-the-art benchmarks in learning performance.
Moreover, the proposed JCC principle validated its advantage by achieving smaller MSE, higher sum rate, and better classification accuracy, compared with classical transceiver configuration schemes.

\appendices

\section{Proof of Lemma \ref{lemma_1}}
\label{proof_of_lemma_1}
By plugging ${\bf{w}}={\bf{w}}_{t+1}$ and ${\bf{w}}'={\bf{w}}_t$ into (\ref{assumption_2}), we have
\begin{align}
	\label{consecutive_fundation_1}
	F({\bf{w}}_{t+1}) \overset{~}{\le} & F({\bf{w}}_t) + \left({\bf{w}}_{t+1} - {\bf{w}}_t \right)^{\rm T}{\nabla F({\bf{w}}_t)} \notag \\
	&+ \frac{L}{2} \|{\bf{w}}_{t+1} - {\bf{w}}_t\|^2 \notag \\
	\overset{(a)}{=}& F({\bf{w}}_t) - \eta \|\nabla F({\bf{w}}_t)\|^2 + \eta {\bf{e}}^{\rm T} {\nabla F({\bf{w}}_t)}\notag \\
	&+ \frac{L\eta^2}{2} \|\nabla F({\bf{w}}_t) - {\bf{e}}\|^2 \notag \\ 
	\overset{(b)}{=}& F({\bf{w}}_t) - \frac{1}{2L} \|\nabla F({\bf{w}}_t)\|^2 + \frac{1}{2L} \|{\bf{e}}\|^2,
\end{align}
where $(a)$ comes from plugging (\ref{actual_model_update}) into the right-hand side of (\ref{consecutive_fundation_1}), and (b) stems from letting $\eta=1/L$.
From (\ref{definition_of_e}), it follows that
\begin{align}
	\label{consecutive_fundation_2}
	\|{\bf{e}}\|^2 \overset{~}{=}& \| a_1({\nabla F({\bf{w}}_t)} - {\bf{g}}_t^f ) + a_2({\nabla F({\bf{w}}_t)}-{\bf{g}}_t^c) \notag \\
	&+ a_1( {\bf{g}}_t^f  - {\hat{\bf{g}}}_t^f ) \|^2 \notag \\
	\overset{(a)}{\le}& 2\| a_1({\nabla F({\bf{w}}_t)} - {\bf{g}}_t^f ) + a_2({\nabla F({\bf{w}}_t)}-{\bf{g}}_t^c) \|^2 \notag \\
	&+ 2{a_1^2}\| {\bf{g}}_t^f - {\hat{\bf{g}}}_t^f\|^2 \notag \\
	\overset{(b)}{\le}& 4{a_1^2} \| {\nabla F({\bf{w}}_t)} - {\bf{g}}_t^f \|^2 + 4{a_2^2} \| {\nabla F({\bf{w}}_t)} - {\bf{g}}_t^c \|^2 \notag \\
	&+ 2{a_1^2}\| {\bf{g}}_t^f  - {\hat{\bf{g}}}_t^f\|^2.
\end{align}
\noindent Here, $(a)$ and $(b)$ come from the Cauchy-Schwarz inequality and triangle inequality, respectively.
Finally, we have (\ref{consecutive_convergence}) by plugging (\ref{consecutive_fundation_2}) into (\ref{consecutive_fundation_1}) and taking the expectation on both sides.
This completes the proof.

\section{Proof of Lemma \ref{lemma_2}}
\label{proof_of_lemma_2}
Based on definitions of ${\nabla F({\bf{w}}_t)}$ and ${\bf{g}}_t^f$, we first bound $\|{\bf{e}}_1\|^2$ as follows:
\begin{align}
	\label{norm_e1}
	\|{\bf{e}}_1\|^2\overset{~}{=}&\|{\nabla F({\bf{w}}_t)} - {\bf{g}}_t^f\|^2 \notag \\ 
	\overset{~}{=}& \left\| \frac{1}{N} \sum\nolimits_{k=1}^K \sum\nolimits_{n \in \mathcal{D}_k} {{\bf{g}}_{t,k,n}} \right. \notag \\ 
	&\left. -\frac{1}{N_f} \sum\nolimits_{k=1}^K \sum\nolimits_{n \in \mathcal{D}_{f,t,k}} {{\bf{g}}_{t,k,n}} \right\|^2 \notag \\
	\overset{(a)}{\le}& \frac{1}{N N_f^2} \left[ \sum\nolimits_{n \in (\cup_k\mathcal{D}_{f,t,k})} { (N_f-N)^2\|{\bf{g}}_{t,k,n} \|^2} \right. \notag \\
	&\left. + \sum\nolimits_{n \in \mathcal{D}/{(\cup_k\mathcal{D}_{f,t,k})} } { N_f^2 \| {\bf{g}}_{t,k,n} \|^2} \right] \notag \\
	\overset{(b)}{\le} & \frac{N_f (N_f-N)^2 \!+\! (N-N_f) N_f^2}{N N_f^2} \left(\xi_1 \!+\! \xi_2 \|\nabla F({\bf{w}}_t)\|^2\right) \notag \\
	\overset{~}{=} & \frac{N-N_f}{N_f}(\xi_1 + \xi_2\|{\nabla F({\bf{w}}_t)}\|^2),
\end{align}
where $(a)$ holds because of the triangle inequality and Cauchy-Schwarz inequality, and $(b)$ comes from Assumption~\ref{assumption_3_text}.
By taking the expectation on both sides of (\ref{norm_e1}), we reach (\ref{bound_of_e1}).

Similarly, $\|{\bf{e}}_2\|^2$ can be bounded as follows:
\begin{align}
	\label{norm_e2}
	\|{\bf{e}}_2\|^2\overset{~}{=}&\|{\nabla F({\bf{w}}_t)} - {\bf{g}}_t^c\|^2 \notag \\ 
	\overset{~}{=}& \| \frac{1}{N} \sum\nolimits_{k=1}^K \sum\nolimits_{n \in \mathcal{D}_k} {{\bf{g}}_{t,k,n}} -\frac{1}{N_c} \sum\nolimits_{n \in \mathcal{D}_{c,t}} {{\bf{g}}_{t,n}} \|^2 \notag \\
	\overset{~}{\le}& \frac{1}{N N_c^2} \left[ \sum\nolimits_{n \in \mathcal{D}_{c,t}} { (N_c\!-\!N)^2\|{\bf{g}}_{t,k,n} \|^2} \right. \notag \\
	&\left. + \sum\nolimits_{n \in \mathcal{D}/{\mathcal{D}_{c,t}} } { N_c^2 \| {\bf{g}}_{t,k,n} \|^2} \right]\notag \\ 
	\overset{~}{\le}& \frac{N-N_c}{N_c}(\xi_1 \!+\! \xi_2\|{\nabla F({\bf{w}}_t)}\|^2).
\end{align}
We are able to obtain (\ref{bound_of_e2}) by taking the expectation on both sides of (\ref{norm_e2}).

Based on ${\bf{g}}_t^f=\sum\nolimits_{k=1}^{K}{\frac{N_{f,k}}{N_f}{\bf{g}}^f_{t,k}}$ and (\ref{actual_aggregated_model}), we rewrite $\|{\bf{e}}_3\|^2$  as follows:
\begin{align}
	\label{norm_e3_1}
	\|{\bf{e}}_3\|^2 =& \|{\bf{g}}_t^f - {\hat{\bf{g}}}_t^f\|^2 \notag \\
	=& \sum \nolimits_{q=1}^Q {\left| \sum \nolimits_{k=1}^K {\zeta_k(g^f_{t,k,q}-{\bar{g}_t})} - {\bar{\sigma}_t}{\bf{b}}_t^{\rm H}{\bf{n}}_{t,q} \right|^2},
\end{align}
\noindent where $\zeta_k = (N_{f,k}/N_f) (1 - {p_{t,f,k}} {\bf{b}}_t^{\rm H} {{\bf{h}}_{t,k}}) $.
Taking the expectation w.r.t. $\{{\bf{n}}_{t,q}\}$ on both sides, we have
\begin{align}
	\label{norm_e3_2}
	\mathbb{E}[ \|{\bf{e}}_3\|^2 ] \overset{~}{=} & \sum \nolimits_{q=1}^Q  {\left| \sum \nolimits_{k=1}^K {\zeta_k (g^f_{t,k,q}-{\bar{g}_t})} \right|^2} \notag \\
	&+ Q{\bar{\sigma}_t^2}{\sigma^2}\|{\bf{b}}_t\|^2 \notag \\ 
	\overset{(a)}{\le} & \sum \nolimits_{q=1}^Q {( \sum \nolimits_{k=1}^K {|\zeta_k|^2} ) (\sum \nolimits_{k=1}^K  {|g^f_{t,k,q}-{\bar{g}_t}|^2} )} \notag \\
	&+ Q{\bar{\sigma}_t^2}{\sigma^2}\|{\bf{b}}_t\|^2 \notag \\
	\overset{(b)}{\le} & 2 ( \sum \nolimits_{k=1}^K \!{|\zeta_k|^2}) ( \sum \nolimits_{k=1}^K  \sum\nolimits_{q=1}^Q \! {| g^f_{t,k,q} |^2} \!+\! KQ | {\bar{g}}_t |^2   ) \notag \\
	&+ Q{\bar{\sigma}_t^2}{\sigma^2}\|{\bf{b}}_t\|^2,
\end{align}
where $(a)$ and $(b)$ are due to the Cauchy-Schwarz inequality.
Based on definitions of ${\bar{g}}_t$ and $\bar{\sigma}_t^2$ in Section~\ref{communication_model}, we bound them as follows:
\begin{align}
	\label{bound_of_mean}
	|{\bar{g}}_t|^2  \overset{(a)}{\le}& \frac{1}{KQ} \sum \nolimits_{k=1}^K \sum \nolimits_{q=1}^Q {(g^f_{t,k,q})^2} \notag \\ 
	\overset{(b)}{=}& \frac{1}{KQ} \sum \nolimits_{k=1}^K {\|{\bf{g}}^f_{t,k}\|^2}, \\
	\label{bound_of_variance}
	\bar{\sigma}_t^2 \overset{~}{\le}& \frac{1}{KQ} \sum\nolimits_{k=1}^K {\sum\nolimits_{q=1}^Q{(g^f_{t,k,q})^2}}\notag \\ 
	\overset{~}{=}& \frac{1}{KQ} \sum \nolimits_{k=1}^K {\|{\bf{g}}^f_{t,k}\|^2},
\end{align}
where $(a)$ comes from the Cauchy-Schwaz inequality, and $(b)$ holds because $\sum \nolimits_{q=1}^Q {|g^f_{t,k,q}|^2} = \|{\bf{g}}^f_{t,k}\|^2$.
Based on (\ref{bound_of_mean}) and (\ref{bound_of_variance}), we further derive (\ref{norm_e3_2}) as
\begin{align}
	\label{norm_e3_3}
	\mathbb{E}[ \|{\bf{e}}_3\|^2 ] \overset{~}{\le}& 4 \left( \sum \nolimits_{k=1}^K {|\zeta_k|^2} \right)\left ( \sum \nolimits_{k=1}^K {\|{\bf{g}}^f_{t,k}\|^2} \right) \notag \\
	&+ \frac{{\sigma^2}\|{\bf{b}}_t\|^2}{K} \sum \nolimits_{k=1}^K {\|{\bf{g}}^f_{t,k}\|^2} \notag \\
	\overset{(a)}{\le}&\frac{4KG^2}{N_f^2} \sum\nolimits_{k=1}^K \!{N_{f,k}^2 | 1 - p_{t,f,k} {\bf{b}}_t^{\rm H} {\bf{h}}_{t,k} |^2} \notag \\
	&+ G^2 \sigma^2 \|{\bf{b}}_t\|^2,
\end{align}
where $(a)$ comes from taking the expectation w.r.t. $\|{\bf{g}}^f_{t,k}\|^2$ on both sides of (\ref{norm_e3_3}) while utilizing Assumption \ref{assumption_3_text}.
This completes the proof.

\section{Proof of Theorem \ref{theorem_1}}
\label{proof_of_theorem_1}
By plugging the three upper bounds in Lemma~\ref{lemma_2} into Lemma~\ref{lemma_1}, we have
\begin{align}
	\label{consecutive_convergence_1}
	&\mathbb{E}\left[ F({\bf{w}}_{t+1}) \right]  \le  \mathbb{E}\left[ F({\bf{w}}_t) \right] + 2\xi_1\frac{N_f(N\!-\!N_f)\!+\!N_c(N\!-\!N_c)}{L(N_f+N_c)^2} \notag \\
	&+ \frac{4KG^2\! \sum\nolimits_{k=1}^K \! {N_{f,k}^2 \! \left| 1 \!\!-\! p_{t,f,k} {\bf{b}}_t^{\rm H} {\bf{h}}_{t,k} \right|^2}\!\!\!+\!\!N_f^2G^2\sigma^2\|{\bf{b}}_t\|^2}{L(N_f\!+\!N_c)^2} \notag \\ 
	&- \left[ \frac{1}{2L} \!-\! 2\xi_2\frac{N_f(N\!-\!N_f)\!+\!N_c(N\!-\!N_c)}{L(N_f\!+\!N_c)^2} \right]\|{\nabla F({\bf{w}}_t)}\|^2.\!\!
\end{align}
Then, we minimize the left-hand side of (\ref{assumption_1}) by plugging in ${\bf{w}}={\bf{w}}^*$,  while minimizing the right-hand side of (\ref{assumption_1}) by setting ${\bf{w}}'={\bf{w}}_t$ and ${\bf{w}}={\bf{w}}_t-\frac{1}{\mu}{\nabla F({\bf{w}}_t)}$.
As a result, we have the following PL inequality~\cite{Ni2022STAR}:
\begin{align}
	\label{pl_inequality}
	\|{\nabla F({\bf{w}}_t)}\|^2 \ge 2\mu\left( F({\bf{w}}_t) - F({\bf{w}}^*) \right).
\end{align}
By plugging (\ref{pl_inequality}) into (\ref{consecutive_convergence_1}), we have
\begin{align}
	\label{consecutive_convergence_2}
	&\mathbb{E}\left[ F({\bf{w}}_{t+1}) \right] \le \mathbb{E}\left[ F({\bf{w}}_t) \right] + \varphi_{t}\left( \{ p_{f,k} \}, {\bf{b}} \right) + \rho_2 \notag \\
	&-\left[ \frac{\mu}{L} - 4\mu\xi_2\frac{N_f(N\!\!-\!\!N_f)\!+\!N_c(N\!\!-\!\!N_c)}{L(N_f\!+\!N_c)^2} \right]( F({\bf{w}}_t) \!\!-\!\! F({\bf{w}}^*) ).
\end{align}
Subtracting $F({\bf{w}}^*)$ and taking the expectation on both sides of (\ref{consecutive_convergence_2}), we have
\begin{align}
	\label{consecutive_convergence_3}
	\mathbb{E}\left[ F({\bf{w}}_{t+1}) - F({\bf{w}}^*) \right] \le& \rho_1\mathbb{E}\left[ F({\bf{w}}_{t}) - F({\bf{w}}^*) \right]  \notag \\
	&+ \rho_2 + \varphi_{t}\left( \{ p_{f,k} \}, {\bf{b}} \right).
\end{align}
Recursively applying (\ref{consecutive_convergence_3}) for $t$ times, it holds that
\begin{align}
	\mathbb{E}\left[ F({\bf{w}}_{t+1}) - F({\bf{w}}^*) \right] \le & \rho_1^t\mathbb{E}\left[ F({\bf{w}}_{1}) - F({\bf{w}}^*) \right] \notag \\
	&+ \rho_2 \frac{1-\rho_1^t}{1-\rho_1} \notag \\
	&+ \sum\nolimits_{i=0}^{t-1} {\rho_1^i\varphi_{t-i}\left( \{ p_{f,k} \}, {\bf{b}} \right)}.
\end{align}
We reach (\ref{convergence_bound}) by setting $t=T$.
This completes the proof.

\section{Proof of Corollary \ref{corollary_new_1}}
\label{proof_corollary_new_1}
When the BS uses all accumulated $\bar{N}_{c,t}=tN_c=t\sum\nolimits_{k=1}^K N_{c,k}$ data samples till the $t$-th round for CL, the CL gradient is calculated by $\bar{{\bf{g}}}^c_t = (1/\bar{N}_{c,t}) \sum\nolimits_{i=1}^{t}\sum\nolimits_{k=1}^{K}\sum\nolimits_{n \in \mathcal{D}_{c,i,k}} {{\bf{g}}_{i,k,n}}$.
Accordingly, the global gradient of the $t$-th round is re-calculated by $\hat{{\bf{g}}}_t=\bar{a}_{1,t}\hat{{\bf{g}}}_t^f + \bar{a}_{2,t}\bar{{\bf{g}}}^c_t$, where $\bar{a}_{1,t}=N_f/(N_f+\bar{N}_{c,t})$ and $\bar{a}_{2,t}=\bar{N}_{c,t}/(N_f+\bar{N}_{c,t})$.
As a result, the gradient error ${\bf{e}}_2$ is rewritten as ${\bf{e}}_{2,t}\!=\!\nabla F({\bf{w}}_t)\!-\!\bar{{\bf{g}}}^c_t$ for the $t$-th round, which proves to be bounded as follows based on (\ref{norm_e2}):
\begin{align}
	\label{new_gradient_error_2}
	\mathbb{E}[\|{\bf{e}}_{2,t}\|^2] \le \frac{N-\bar{N}_{c,t}}{\bar{N}_{c,t}}(\xi_1+\xi_2 \|\nabla F({\bf{w}}_t)\|^2), \forall t.
\end{align}
By plugging $\bar{a}_{1,t}$, $\bar{a}_{2,t}$, and (\ref{new_gradient_error_2}) into Lemma~\ref{lemma_1}, we have (\ref{optimali_gap_all_samples}) after applying the same mathematical derivation in Appendix~\ref{proof_of_theorem_1}. This completes the proof.

\section{Proof of Corollary \ref{corollary_new_2}}
\label{proof_corollary_new_2}
By substituting ${\bf{w}}={\bf{w}}_{t+1}$ and ${\bf{w}}'={\bf{w}}_t$ into (\ref{assumption_2}), we have
\begin{align}
	F({\bf{w}}_{t+1}) -  F({\bf{w}}_t)  \overset{~}{\le}&  -\eta_t \hat{\bf{g}}_t^{\rm T}{\nabla F({\bf{w}}_t)}  +  \frac{L\eta^2}{2} \|\hat{\bf{g}}_t\|^2 \notag \\ 
	\overset{~}{=}& \frac{\eta_t}{2}\left(L\eta_t-1\right)\|\hat{\bf{g}}_t\|^2  -  \frac{\eta_t}{2} \|\nabla F({\bf{w}}_t)\|^2 \notag \\
	&+ \frac{\eta_t}{2} \|{\bf{e}}_t\|^2 \notag \\
	\overset{(a)}{\le}& - \frac{\eta_t}{2} \|\nabla F({\bf{w}}_t)\|^2 + \frac{\eta_t}{2} \|{\bf{e}}_t\|^2,
\end{align}
where $(a)$ is because $\tau \ge \Lambda L$ so that $\eta_t = \frac{\Lambda}{t+\tau} \le \frac{1}{L}$.
By applying the PL inequality in (\ref{pl_inequality}) to the right-hand side and taking the expectation on both sides, while plugging $\eta_t\!=\!\frac{\Lambda}{t+\tau}$ and $\theta_t\!=\!\max\{\frac{\Lambda (t+\tau) \mathbb{E}[\|{\bf{e}}_t\|^2]}{2 (\Lambda \mu - 1)},\mathbb{E}[F({\bf{w}}_t)\!-\!F({\bf{w}}^*)](t\!+\!\tau)\}$, we obtain
\begin{align}
	\mathbb{E}[F({\bf{w}}_{t+1})-F({\bf{w}}^*)] \overset{~}{\le}& \frac{t+\tau-1}{(t+\tau)^2}\theta_t \notag \\ 
	&- \frac{\Lambda\mu\!-\!1}{(t\!+\!\tau)^2} \left[ \theta_t - \frac{\Lambda(t+\tau)\mathbb{E}[\|{\bf{e}}_t\|^2]}{2(\Lambda\mu-1)} \right] \notag \\ \overset{(b)}{\le}& \frac{\theta_t}{(t+\tau+1)},
\end{align}
where $(b)$ is due to the definition of $\theta_t$ and $\frac{t+\tau-1}{(t+\tau)^2}\le\frac{1}{t+\tau+1}$.
We have (\ref{optimality_gap_decreasing_learning_rate}) by setting $t=T$.
	
With sufficient training rounds and careful optimization of transceivers, one can suppose that $\lim_{t \rightarrow \infty} \mathbb{E}[\|{\bf{e}}_t\|^2] =0$.
This implies that there is a $\tilde{t}>0$ such that $\frac{\Lambda (t+\tau) \mathbb{E}[\|{\bf{e}}_t\|^2]}{2 (\Lambda \mu - 1)} \le \mathbb{E}[F({\bf{w}}_t)-F({\bf{w}}^*)](t+\tau), \forall t \ge \tilde{t}$.
Therefore, one can verify that
\begin{align}
	\mathbb{E}[F({\bf{w}}_{t+1})-F({\bf{w}}^*)]  \le& \frac{\mathbb{E}[F({\bf{w}}_{t})-F({\bf{w}}^*)](t+\tau)}{t+\tau+1}\notag \\ 
	\le&\frac{\mathbb{E}[F({\bf{w}}_{\tilde{t}})- F({\bf{w}}^*)](\tilde{t}+\tau)}{t+\tau+1}, \forall t \ge \tilde{t}.
\end{align}
Through replacing $t$ with $T$, we have $\mathbb{E}[F({\bf{w}}_{T+1}\!)\!-\!\!F({\bf{w}}^*)] \!\! \rightarrow \! 0$ as $T \!\! \rightarrow \! \infty$. 
This completes the proof.

\section{Proof of Theorem \ref{theorem_2}}
\label{proof_of_theorem_2}
For FL, the BS only aggregates local gradients, which implies $\hat{\bf{g}}_t=\hat{\bf{g}}_t^f$.
The global gradient error reduces to ${\bf{e}}_{FL} = \tilde{\bf{e}}_1 + \tilde{\bf{e}}_3$, where $\tilde{\bf{e}}_1=\nabla F({\bf{w}}_t)-{\bf{g}}_t^f$ and $\tilde{\bf{e}}_3={\bf{g}}_t^f-\hat{\bf{g}}_t^f$.
Taking the expectation on both sides of (\ref{consecutive_fundation_1}), we have
\begin{align}
	\label{FL_consecutive_fundation_1}
	\mathbb{E}[F({\bf{w}}_{t+1})] \overset{~}{\le}& \mathbb{E}[F({\bf{w}}_t)] - \frac{1}{2L} \mathbb{E}[\|\nabla F({\bf{w}}_t)\|^2] \notag \\
	&+ \frac{4}{L} \mathbb{E}[\|\tilde{\bf{e}}_{1}\|^2] +  \frac{1}{L} \mathbb{E}[\|\tilde{\bf{e}}_{3}\|^2].
\end{align}
By separately substituting $N_f$ with $N_f+N_c$ and $N_{f,k}$ with $N_{f,k}+N_{c,k}$ into Lemma~\ref{lemma_2}, we have $\mathbb{E}[\|\tilde{\bf{e}}_1\|^2]$ and $\mathbb{E}[\|\tilde{\bf{e}}_3\|^2]$ are bounded, respectively, by
\begin{align}
	\label{FL_gradient_bound_1}
	\mathbb{E}[\|\tilde{\bf{e}}_1\|^2] \le& \frac{N-(N_f+N_c)}{N_f+N_c}(\xi_1 + \xi_2\|{\nabla F({\bf{w}}_t)}\|^2), \\
	\label{FL_gradient_bound_3}
	\mathbb{E}[\|\tilde{\bf{e}}_3\|^2] \le& \frac{4KG^2}{(N_f+N_c)^2}\! \sum\limits_{k=1}^K \! {(N_{f,k}+N_{c,k})^2 \! \left| 1\! -\! p_{t,f,k} {\bf{b}}_t^{\rm H} {\bf{h}}_{t,k} \right|^2} \notag \\ 
	&+ G^2 \sigma^2 \|{\bf{b}}_t\|^2.
\end{align}
After plugging (\ref{FL_gradient_bound_1}) and (\ref{FL_gradient_bound_3}) into (\ref{FL_consecutive_fundation_1}) while subtracting $F({\bf{w}}^*)$ on both sides, we have
\begin{align}
	\label{FL_consecutive_fundation_2}
	\mathbb{E}\left[ F({\bf{w}}_{t+1}) - F({\bf{w}}^*) \right] \le& \tilde{\rho}_1 \mathbb{E}\left[ F({\bf{w}}_t) - F({\bf{w}}^*) \right] + \tilde{\rho}_2 \notag \\
	&+\tilde{\varphi}_{t}\left( \{ p_{f,k} \}, {\bf{b}} \right).
\end{align}
Recursively applying (\ref{FL_consecutive_fundation_2}) for $t$ times and letting $t=T$, (\ref{convergence_bound_FL}) can be obtained.

For CL, the BS calculates the global gradient based on the randomly selected $N_f+N_c$ data samples in $\mathcal{D}_{c,t}$, i.e., $\hat{\bf{g}}_t=\frac{1}{N_f+N_c} \sum_{n \in \mathcal{D}_{c,t}} {\bf{g}}_{t,n}$.
The global gradient error becomes ${\bf{e}}_{CL}=\nabla F({\bf{w}}_t) - \hat{\bf{g}}_t$.
Similarly, taking the expectation on both sides of (\ref{consecutive_fundation_1}), we have
\begin{align}
	\label{CL_consecutive_fundation_1}
	\mathbb{E}[F({\bf{w}}_{t+1})] \overset{~}{\le}& \mathbb{E}[F({\bf{w}}_t)] - \frac{1}{2L} \mathbb{E}[\|\nabla F({\bf{w}}_t)\|^2] \notag \\
	&+ \frac{1}{2L} \mathbb{E}[\|{\bf{e}}_{CL}\|^2].
\end{align}
By substituting $N_c$ in (\ref{bound_of_e2}) with $N_f+N_c$, we have $\|{\bf{e}}_{CL}\|^2 \le \frac{N-(N_f+N_c)}{N_f+N_c}(\xi_1+\xi_2\|\nabla F({\bf{w}}_t)\|^2)$.
As a result, it holds that
\begin{align}
	\label{CL_consecutive_fundation_2}
	\mathbb{E}\left[ F({\bf{w}}_{t+1}) - F({\bf{w}}^*) \right] \le \hat{\rho}_1 \mathbb{E}\left[ F({\bf{w}}_t) - F({\bf{w}}^*) \right] + \hat{\rho}_2.
\end{align}
Recursively applying (\ref{CL_consecutive_fundation_2}) for $t$ times and setting $t=T$, (\ref{convergence_bound_CL}) can be obtained.

To reveal the relation between SemiFL, FL, and CL, one can find $\tilde{\rho}_1 \ge \rho_1$, $\tilde{\rho}_2 \ge \rho_2$, and $\tilde{\varphi}_{t}\left( \{ p_{f,k} \}, {\bf{b}} \right) \ge \varphi_{t}\left( \{ p_{f,k} \}, {\bf{b}} \right)$, since $N \gg N_f+N_c$ and $\frac{(N_{f,k}+N_{c,k})^2}{(N_f+N_c)^2} \ge \frac{N_{f,k}^2}{(N_f+N_c)^2}$.
Hence, $\psi_T^{\rm SemiFL}( \{ p_{f,k} \}, {\bf{b}} ) \le \psi_T^{\rm FL}( \{ p_{f,k} \}, {\bf{b}} )$ holds.
By separately rewriting $\rho_1$ and $\rho_2$ as
\begin{align}
	&\rho_1 = \hat{\rho}_1 + \frac{3 \mu \xi_2 (N_f+N_c)[N-(N_f+N_c)]+8 \mu \xi_2 N_f N_c}{L(N_c+N_f)^2}, \\
	&\rho_2 = \hat{\rho}_2 + \frac{\xi_1(N_f+N_c)[N-(N_f+N_c)]+4\xi_1 N_c N_f}{L(N_c+N_f)^2},
\end{align}
we find that $\rho_1 \ge \hat{\rho}_1$ and $\rho_2 \ge \hat{\rho}_2$.
Since $\varphi_{t}( \{ p_{f,k} \}, {\bf{b}} ) \ge 0$, we have $\psi_T^{\rm CL}( \{ p_{f,k} \}, {\bf{b}} ) \le \psi_T^{\rm SemiFL}( \{ p_{f,k} \}, {\bf{b}} )$.
This completes the proof.

\section{Proof of Lemma \ref{lemma_3}}
\label{proof_of_lemma_3}
When the BS has a single antenna, the aggregation beamformer $\bf{b}$ degrades to a scalar $b\in\mathbb{C}$. The Lagrange function of problem (\ref{p_5}) is given by
\begin{align}
	\label{Lagrange_function_b}
	\mathcal{L}(b,\lambda)=&b^{\rm H}(\omega_0+\lambda \omega_1)b-2{\mathop{\rm Re}\nolimits} \{ b^{\rm H}(\hat{h}_0+\lambda\hat{h}_1) \} \notag \\
	&+\lambda(\iota-\epsilon),
\end{align}
where $\lambda \ge 0$ is the Lagrange multiplier. Then, KKT conditions are given by
\begin{subnumcases}
	{\label{KKT_b}}
	\label{gradient_Lagrange_function_b}
	\frac{\partial \mathcal{L}(b,\lambda)}{\partial b}=0, \\
	\lambda \ge 0, \\
	\label{complementary_slackness_b}
	\lambda(b^{\rm H}\omega_1b-2{\mathop{\rm Re}\nolimits} \{ b^{\rm H}\hat{h}_1 \}+\iota-\epsilon)=0.
\end{subnumcases}
By solving (\ref{gradient_Lagrange_function_b}), we obtain
\begin{align}
	\label{b_with_multiplier}
	b=\frac{1}{\omega_0\!+\!\lambda \omega_1} \sum\nolimits_{k=1}^K{\left[\frac{4KN_{f,k}^2}{(N_f+N_c)^2}\!+\!\frac{N_{f,k}^2}{N_f^2}\lambda\right]p_{f,k}h_k}.
\end{align}
We plug (\ref{b_with_multiplier}) into equation $b^{\rm H}\omega_1b-2{\mathop{\rm Re}\nolimits} \{ b^{\rm H}\hat{h}_1 \}+\iota-\epsilon = 0$ and solve it to obtain $\lambda$ as
\begin{align}
	\label{solution_lambda}
	\lambda=-\frac{\omega_0}{\omega_1}+\frac{|\omega_0\hat{h}_1-\omega_1\hat{h}_0|}{\omega_1\sqrt{|\hat{h}_1|^2-\omega_1(\iota-\epsilon)}}.
\end{align}
Plugging (\ref{solution_lambda}) into (\ref{b_with_multiplier}), we reach (\ref{b_solution_single_antenna}).
This completes the proof.

\section{Proof of Lemma \ref{lemma_4}}
\label{proof_of_lemma_4}
The Lagrange function of problem (\ref{p_10}) is given by
\begin{align}
	\label{Lagrange_function_f}
	\mathcal{L}({\bf{f}}_k,\nu_k,\lambda_1,\lambda_2)=\lambda_1 g({{\bf{f}}}_k|{{\bf{f}}}^{(n)}_k)+ (1-\lambda_1+\lambda_2)\nu_k,
\end{align}
where $\lambda_1$ and $\lambda_2$ are Lagrange multipliers. 
Then, KKT conditions are given by
\begin{subnumcases}
	{\label{KKT_f}}
	\label{gradient_Lagrange_function_f}
	\frac{\partial \mathcal{L}({\bf{f}}_k,\nu_k,\lambda_1,\lambda_2)}{\partial {\bf{f}}_k}={\bf{0}}, \\
	\label{gradient_Lagrange_function_nu}
	\frac{\partial \mathcal{L}({\bf{f}}_k,\nu_k,\lambda_1,\lambda_2)}{\partial \nu_k}=0, \\
	\lambda_1 \ge 0, \lambda_2 \ge 0,\\
	\label{complementary_slackness_f}
	\lambda_1(g({{\bf{f}}}_k|{{\bf{f}}}^{(n)}_k)-\nu_k)=0, \\
	\label{complementary_slackness_nu}
	\lambda_2\nu_k=0.
\end{subnumcases}
By solving (\ref{gradient_Lagrange_function_f}) and (\ref{gradient_Lagrange_function_nu}), we have
\begin{align}
	\label{solution_f_with_multplier}
	&{\bf{f}}_k=\frac{1}{\lambda_1}{\bf{M}}^{-1}_k({\bf{M}}_k-{\bf{A}}_{2,k}){\bf{f}}^{(n)}_k, \\
	\label{relation_between_multipliers}
	&\lambda_2-\lambda_1+1=0.
\end{align}

If $\lambda_2>0$,  we have $\nu_k=0$ based on (\ref{complementary_slackness_nu}). 
However, according to (\ref{complementary_slackness_f}), one can prove that there is no solution for $\lambda_1$ by solving equation $g({{\bf{f}}}_k|{{\bf{f}}}^{(n)}_k)\!=\!0$ when $\nu_k\!=\!0$.
As such, we have $\lambda_2\!=\!0$.
In terms of (\ref{solution_f_with_multplier}), $\lambda_1$ can not be $0$, which implies $\lambda_1\!>\!0$.
By plugging $\lambda_2\!=\!0$ into (\ref{relation_between_multipliers}), we have $\lambda_1\!=\!1$ and reach (\ref{f_k_solution}).
Considering the complementary slackness condition (\ref{complementary_slackness_f}), we have $g({{\bf{f}}}_k|{{\bf{f}}}^{(n)}_k)\!-\!\nu_k\!=\!0$. 
By plugging (\ref{f_k_solution}) into $g({{\bf{f}}}_k|{{\bf{f}}}^{(n)}_k)$, we reach (\ref{nu_k_solution}).
This completes the proof.

\vspace{-0.4 cm}
\bibliographystyle{IEEEtran}
\bibliography{reference}

\end{document}